\documentclass[iop]{emulateapj}
\usepackage{color} \usepackage{enumerate}
\usepackage{graphicx,subfigure} \usepackage{fancyheadings}
\usepackage{longtable} \usepackage{rotating}
\usepackage{epsfig} \usepackage[fleqn]{amsmath}
\usepackage{mathtools} \usepackage{multirow}
\usepackage{ctable}
\usepackage{mathrsfs}
\usepackage{tikz}
\usepackage{color,soul}

\begin{document}

\newcommand{\dg}{$^{\circ}$} 
\newcommand{\textoverline}[1]{$\overline{\mbox{#1}}$}
\newcommand{\sex}{{\it SExtractor}}
\newcommand{\Msol}{M$_{\odot}$}
\newcommand{\Lsol}{L$_{\odot}$}
\newcommand{\col}{$\colon$}
\newcommand{\gsim}{$\gtrsim$}
\newcommand{\nod}{\nodata}
\newcommand{\lsim}{$\lesssim$}
\newcommand{\snr}{signal-to-noise ratio}
\newcommand{\rsrang}{$0.9<z<1.4$}
\newcommand{\nuvu}{(NUV--U$^{\prime})$}
\newcommand{\ha}{H$\alpha$}
\newcommand{\fuvu}{(FUV--U$^{\prime})$}
\newcommand{\fesc}{$f_{esc}$}
\newcommand{\ifrac}{$x_{\mbox{\tiny \ion{H\!\!}{2}}}$}
\newcommand{\rhouv}{$\rho_{UV}$}
\newcommand{\dittotikz}{%
    \tikz{
        \draw [line width=0.12ex] (-0.2ex,0) -- +(0,0.8ex)
            (0.2ex,0) -- +(0,0.8ex);
        \draw [line width=0.08ex] (-0.6ex,0.4ex) -- +(-1.5em,0)
            (0.6ex,0.4ex) -- +(1.5em,0);
    }%
}
\title{\bf The Lyman Continuum Escape Fraction of Low-Mass Star-Forming Galaxies at \lowercase{z}$\sim$1}

\author{Michael J. Rutkowski\altaffilmark{1,$\dag$}, Claudia Scarlata\altaffilmark{1}, Francesco Haardt\altaffilmark{2}, Brian Siana\altaffilmark{3}, Alaina Henry\altaffilmark{4,$\ddag$}, Marc Rafelski\altaffilmark{4,$\ddag$},  Matthew Hayes\altaffilmark{5}, Mara Salvato\altaffilmark{6}
  Anthony J. Pahl\altaffilmark{1}, Vihang Mehta\altaffilmark{1}, Melanie Beck\altaffilmark{1}, Matthew Malkan\altaffilmark{7}, and Harry I. Teplitz\altaffilmark{8}}
\affil{
\altaffilmark{1}{Minnesota Institute for Astrophysics, University of Minnesota, 116 Church St. SE, Minneapolis, MN 55455, USA}\\
\altaffilmark{2}{Dipartimento di Scienza e alta Tecnologia, Universit\`{a} dell'Insubria, via Valleggio 11, 22100 Como, Italy}\\
\altaffilmark{3}{Department of Physics and Astronomy, University of California, Riverside, CA 92521, USA}\\
\altaffilmark{4}{Astrophysics Science Division, Goddard Space Flight Center, Code 665, Greenbelt, MD 20771, USA}\\
\altaffilmark{5}{Department of Astronomy, Oskar Klein Centre, Stockholm University, AlbaNova University Centre, SE-106 91 Stockholm, Sweden}\\
\altaffilmark{6}{Max Planck Institut f\"{u}r Plasma Physik and Excellence Cluster, 85748 Garching, Germany}\\
\altaffilmark{7}{Astronomy Division, University of California, Los Angeles, CA 90095-1562 USA}\\
\altaffilmark{8}{Infrared Processing and Analysis Center, MS 100-22, Caltech, Pasadena, CA 91125, USA}\\
}
\altaffiltext{$\dag$}{email:\href{mailto:rutkowsk@astro.umn.edu}{rutkowsk@astro.umn.edu}}
\altaffiltext{$\ddag$}{NASA Post-doctoral Program Fellow}
\begin{abstract} 

To date no direct detection of Lyman continuum emission has been
measured for intermediate--redshift ($z\sim1$) star--forming galaxies .
We combine HST grism spectroscopy with GALEX UV and ground--based
optical imaging to extend the search for escaping Lyman continuum to a
large ($\sim$600) sample of $z\sim1$ low--mass
($\mbox{log(\textoverline{M})}\simeq$9.3\Msol), moderately star--forming
($\overline{\Psi}$\lsim$10$\Msol\,yr$^{-1}$) galaxies selected initially
on H$\alpha$ emission. The characteristic escape fraction of LyC from
SFGs that populate this parameter space remains weakly constrained by
previous surveys, but these faint (sub--$L_{\star}$) SFGs are assumed to
play a significant role in the reionization of neutral hydrogen in the
intergalactic medium (IGM) at high redshift $z>6$. We do not make an
unambiguous detection of escaping LyC radiation from this $z\sim1$
sample, individual non--detections to constrain the absolute Lyman
continuum escape fraction, \fesc$<2.1\%$ (3$\sigma$).  We measure an
upper limit of \fesc$<9.6\%$ from a sample of SFGs selected on high
H$\alpha$ equivalent width (EW$>200\mbox{\AA}$), which are thought to be
close analogs of high redshift sources of reionization.  For reference,
we also present an emissivity--weighted escape fraction which is useful
for measuring the general contribution SFGs to the ionizing UV
background. In the discussion, we consider the implications of these
intermediate redshift constraints for the reionization of hydrogen in
the intergalactic medium at high ($z>6$) redshift.  If we assume our
$z\sim1$ SFGs, for which we measure this emissivity-weighted \fesc, ~are
analogs to the high redshift sources of reionization, we find it is
difficult to reconcile reionization by faint (M$_{UV}\lesssim-13$) SFGs
with a low escape fraction (\fesc$<3\%$), with constraints from
independent high redshift observations. If \fesc~evolves with redshift,
reionization by SFGs may be consistent with observations from Planck.

\end{abstract}

\section{Introduction}\label{sec:intro}  

Hydrogen in the inter--galactic medium (IGM) is ionized by the far--UV background of
Lyman continuum photons (LyC, $\lambda<912\mbox{\AA}$), in a process (called reionization) that began 
at $z>10$ (Planck Collaboration 2015) and was complete by $z\sim6$ 
\citep{Djorgovski01,Becker01,Fan02,Siana05}. Star-forming
galaxies (SFGs), quasars (QSOs), and active galactic nuclei (AGN) are
considered the most likely sources of this emission (see the review by
Loeb and Barkana 2001). However, at  $z$\lsim3 QSO alone are
sufficient to mantain the ionized state of the IGM (Haardt and Madau 1996, 2012; Rauch et
al. 1997).  At $z$\gsim3, the space density of QSOs
declines significantly \citep{Osmer82,Faucher08,Siana05}, implying an
alternative source --- presumably SFGs, and possibly
low-luminosity AGN (Giallongo et al. 2015)---must ultimately have
completed reionization. This assumption must be confirmed
observationally, or additional exotic sources of ionizing radiation must
be invoked to explain the onset of reionization --- e.g., dark matter
annihilation in PopIII stars, accretion shocks in massive halos, or mini
quasars \citep[e.g., ][respectively]{Schleicher09,Dopita11,Zaroubi07}.

If galaxies are to ionize hydrogen in the IGM, then it is necessary that
these galaxies must 1) produce sufficient LyC photons in star--forming
regions and 2) these photons must escape the emitting galaxy's interstellar medium (ISM).
Although the first requirement will be met if star forming galaxies
exist in sufficient numbers (Robertson et al. 2015,
Dressler et al. 2015), the second requires that the column density of
neutral hydrogen and dust in the ISM be relatively low.

Because the universe is opaque to LyC radiation at $z$\gsim 4 (Inoue \&
Iwata 2008, Inoue et al.~2014), direct measurements of the escape of ionizing photons,
\fesc, can only be made for low to intermediate redshift SFGs. Thus, our
understanding of the galaxies that initiated  reionization at $z > 6$
must be informed by the study of their lower redshift analogs.  The
measurement of \fesc~for low redshift SFGs has proven difficult, though.
In the local universe, Tol--1247--232 (Leitet et al. 2013) and J0921+4509
(Borthakur et al. 2014) are observed to emit at LyC wavelengths, with
the absolute escape fraction of ionizing photons from young, hot stars
less than $\sim$5\%. A third galaxy, Haro 11, is also likely a LyC
emitter but debated (Bergvall et al.\,2006; Leitet et al.\,2011; cf.
Grimes et al. 2007).

Despite significant efforts with space--based observatories at $z\sim1$,
where the mean IGM opacity to LyC is still less than $\sim50$\%, no
direct detection of escaping LyC from SFGs has been observed using the
Hopkins UV Telescope (Leitherer et al., 1995), the HST Solar Blind Channel (Malkan,
Webb, and Konopacky 2003; Bridge et al.~2010; Siana et al. 2007, 2010)
or GALEX (Cowie, Barger, and Trouille 2009; hereafter CBT09). At higher
redshift ($z\sim3$) the statistics appear to improve --- Iwata et al.
(2009), Nestor et al.~(2011, 2013), Mostardi et al. (2013) have
reported \fesc$\simeq$5---7\%. Yet, claims of direct detection of LyC
emission from $z\sim3$ SFGs remain tentative.  HST imaging of LyC
candidates selected from ground--based surveys often finds these galaxies
to be contaminated by (non--ionizing) UV emission from low-redshift
interlopers (Siana et al.~2015).   Recently,
Mostardi et al. (2015) confirmed 1 (of 16) candidate as a LyC emitter
combining ground- and space- based imaging and spectroscopy and
determined \fesc\, \lsim 15\%, marginally consistent with previous
measurements at similar redshifts. 


Recently, simulations have demonstrated that metal--poor, low--mass
galaxies undergoing a strong episode of star--formation may have a high
escape fraction  \citep{Wise14}. In these low mass galaxies, supernova
feedback may drive outflows, which clear sight--lines for LyC
escape (Fujita et al. 2003; Razoumov and Sommer--Larsen 2006).
Nevertheless, LyC studies require extremely deep observations because
the non--ionizing to ionizing flux (f$_{\nu}$) ratio is of the order $\sim3-10$,
depending on model assumptions. Previous studies at high 
redshifts focused on relatively massive galaxies (e.g.,
M\gsim10$^{9.5}$\Msol, Siana et al. 2010) with high UV luminosities. Here, we use a combination of
GALEX and HST archival data, to study the ionizing emissivity in a large
sample of SFGs selected at $z\sim 1$ via their \ha\ emission line.  The
average stellar mass for our full sample is \textoverline{M}$
=10^{9.3}$\Msol, almost an order of magnitude lower than considered in
previous studies of \fesc~for $z\sim1$ SFGs massive than previous
studies. The size of the sample allows us to explore the dependency of
the LyC escape on various galaxy properties: H$\alpha$ equivalent width,
orientation with respect to the observer, and stellar mass. 

The paper is organized as follows. Section~2 presents the  HST and GALEX
observations, and the reduction of the data. Section~3 we discuss the
steps involved in the selection of our robust  sample of isolated
galaxies, and define the sub--samples used in the analysis of the LyC
escape fraction. In Section 4 we perform a stacking analysis and compute
upper limits to \fesc. Finally, we discuss the implications of these
results for reionization in Section 5.

Throughout, we assume a $\Lambda$CDM cosmology with $\Omega_m=0.27$,
$\Omega_{\Lambda}$=0.73, and H$_{0}=70 $km s$^{-1}$ Mpc$^{-ˆ'1}$ (Komatsu
et al. 2011), and quote all fluxes on the AB magnitude system (Oke and
Gunn 1983).

\section{Observations}
In this work we study the escape fraction of ionizing radiation from low--mass galaxies at $z\sim1$.
The sample is selected using archival IR grism spectroscopy over four deep fields --COSMOS, 
GOODS--North and South, and AEGIS-- obtained as part of the 3DHST (PI: P. van Dokkum) and AGHAST (PI: B.
Weiner) surveys. For the measurement of escaping LyC radiation we exploit  the deep GALEX UV
imaging (m$\simeq$27 mag) available in these fields.  Here, we discuss
the pertinent details associated with the reduction of these data.


\subsection{HST Data Reduction}\label{subsec:thedata}

We performed an independent reduction of the 3DHST WFC3/IR grism (G141)
and direct (F140W) images obtained from the Milkulski Archive for Space
Telescopes (MAST) archive, using a pipeline of processing software
originally developed for the {\it WFC3 IR Spectroscopic Parallel} survey
(WISP; PI: Malkan).  For a full review of the pipeline, we refer the
reader to Atek et al. (2010). Here we summarize key reduction stages in
the processing of these 3DHST data, with specific details provided when
modifications to the original (WISP) pipeline were required.

The 3DHST program observed $\sim$500 sq. arcminutes comprised of
$\sim28$ unique pointings in each four deep fields. Each pointing was
uniformly observed with an exposure time of 800 (F140W, direct image)
and 4700 (G141, dispersed image) seconds.  Each pointing consists of
individual exposures obtained using a four--point dither pattern to
improve the sampling of the (under-sampled) WFC3 IR point spread function
and the rejection of cosmic rays and bad pixels. We used Astrodrizzle
implemented via DrizzlePac \citep{Gonzaga12} to produce drizzled direct
image ``mosaics'' for each of the unique pointings. We adopted a final pixel fraction
parameter, pixfrac = 0.8, and pixel scale direct images with a scale of
0$\farcs$064 per pixel (Nyquist sampled) following Brammer et al.
(2012).   Sky subtraction of the direct images was performed ``on the fly'' as in the WISP pipeline.  A master sky background developed by Atek et
al. was applied for the background subtraction in individual G141 grism
images.  As noted in Brammer et al. (2012), grism data associated with
nine pointings in GOODS--N suffer from a strong sky gradients. These data
can be corrected with higher--order background sky models (cf. Brammer et
al. 2012). We elected to remove the fields from further reduction in the
pipeline because such extreme sky gradients affect only $\sim$10\% of
all 3DHST fields.

We first measured positions and fluxes of sources for each direct image mosaic
using Source Extractor \citep[\sex;][]{Bertins96}. We then used the aXe
software to produce input object lists of source coordinates for objects
in the direct images, and projected these sources to guide the removal
of contaminants and the extraction of object spectra from individual
grism images.  Individual extracted spectra for each G141 image were
then combined using aXeDrizzle tasks to produce a catalog of 1- and
2-dimensional spectra for each unique pointing. 

\subsection{GALEX Images} At $z\sim 1$, ionizing and non--ionizing
radiation are probed by  GALEX FUV and NUV filters, respectively (see,
CBT09). We searched the MAST Archive\footnotetext{maintained by the
Space Telescope Science Institute and available online at
http://galex.stsci.edu/GR6/.} for GALEX primary guest investigator
programs that observed the 3DHST fields for more than 10ks in the FUV
filter.  Due to the total number of observations for each deep field
with GALEX, these data have been prepared as individual coadded stacks.
From the Archive, we downloaded these high--level, 1.25 deg$^2$ science
mosaic frames  (in units of counts s$^{-1}$), and exposure map frames
(``rrhr" frames, which define the effective exposure time per pixel and
account for detector variations). For each field we median combined all
available exposures, using the exposure map frames as weight images. The
total effective exposure time in the FUV range in the 3DHST fields
between 100 and 200ks, because of the different number of programs
observing each field. We converted the mosaics from detector units of
(counts s$^{-1}$) to flux density units (erg s$^{-1}$ cm$^{-2}$
Hz$^{-1}$), assuming the photometric zero points of m$_{FUV,0}$=18.82
and m$_{NUV,0}$=20.08, measured in Cowie et al. (2009).

We computed the depth of each deep image by placing 10$^4$ apertures
over the field of view and measuring the total flux within a $\sim6$\farcs
diameter. This diameter corresponds to a radius $0.673\times\!FWHM_{PSF}$ (where
FWHM$_{FUV}$=4\farcs5, see Martin et al. 2003; Hammer et al. 2010), and
optimizes the Signal--to--Noise (S/N) ratio, under the assumption of a Gaussian
PSF. In Table 1 we present the 3$\sigma$ magnitude limit in the GALEX FUV
images, using the standard deviation of the Gaussian function that best
fits the aperture flux distribution. 

\section{Identification of \rsrang~emission line galaxies}\label{sec:selection}

 For each unique pointing in the 3DHST fields we produced a catalog of
 spectra which were then inspected by two co--authors independently to
 identify emission line galaxies (ELGs). At the resolution and typical
 S/N of the 3DHST grism spectra only bright emission lines are observed
 for $z\sim 1$ SFGs (e.g., [OIII] and \ha). When multiple emission lines
 are present, the line identification---which is then used to measure an
 accurate grism spectroscopic redshift---is straightforward. When only a
 single emission is observed, the user defaulted the classification of
 the line to H$\alpha$, unless the line profile (asymmetric blue
 profile, in a compact source) was suggestive of the [OIII] and H$\beta$
 emission line complex. To ensure robust spectroscopic redshifts
 measured from the grism spectra, we retained in the catalog all SFGs
 that met the following two requirements. First, we consider only
 galaxies for which the independent line--identifications agreed on the
 designation of the line (see Ross et al. 2015 for further details as
 they pertain to the WISP data reduction pipeline). Secondly, we
 included in the catalog only those sources for which photometric and
 spectroscopic redshifts agreed
 ($\Delta(z)=\frac{z_{spec}-z_{phot}}{1+z_{phot}}\lesssim5\%$). The reference photometric
 redshift, $z_{phot}$, used here for comparison with our measured grism
 redshift for each galaxy was drawn from the public photometric redshift
 catalogs\footnote{available online from
 http://3dhst.research.yale.edu/Data.php} produced by the 3DHST team
 (Skelton et al. 2014).

Our spectroscopic redshifts measurements agreed with the photometric
redshift for approximately 50\% of all ELGs.  This fraction improves to
70\% when the photometric redshift catalogs produced by the 3DHST
collaboration are used to correct for mis--classified single emission
lines.  Yet, approximately $\sim30\%$ of all ELGs were measured to have
spectroscopic redshifts which disagreed with the photometric redshift.
We discuss the likely causes for the remaining 30\% of SFGs with
discrepant redshifts in the Appendix.  We adopt a very conservative
strategy in grism spectroscopic redshift selection and exclude all SFGs
with discrepant spectroscopic redshifts that could not be corrected for
the mis--classification of the emission line.  With this cut, we produced
our initial catalog of $\sim$1400 SFGs at \rsrang.

%

\section{Sample Selection}\label{subsec:galids}

When measuring the ionizing radiation below 912\AA, particular care has
to be taken in selecting star--forming galaxies at the proper redshift.
First, we remove AGN from the sample, which we will consider
independently in a future publication.  We used public photometric
redshift catalogs of AGN candidates identified by X--ray and optical--IR
matched surveys of the AEGIS (Laird et al. 2009), GOODS--South and North
(Alexander et al. 2003), and COSMOS (Salvato et al. 2011) fields.
Secondly, we selected star--forming galaxies at $0.9<z<1.4$.  The lower
limit in redshift is fixed by the transmission curve of the GALEX FUV
filter.  The transmission drops below 5\% at $\lambda> 1787$\AA
(Morrissey et al. 2007), ensuring that for SFGs at $z>0.9$ the GALEX FUV
photometry is not significantly affected by non--ionizing emission (see
CBT09). In addition, we remove all SFGs with H$\alpha$ equivalent width
measured to be less than 40\AA, a threshold below which the G141 grism
data are likely to be significantly incomplete ($<80\%$; see Colbert et
al. 2013).

We also apply multiple selection criteria to select for isolated SFGs. 
This is in part necessary due to the coarse GALEX UV spatial resolution.
The FUV GALEX PSF is $\sim$4$\farcs$5, much larger than the spatial
resolution reached with HST. Consequently, lower redshift galaxies in
close spatial proximity to the ELGs of interest may irreducibly
contaminate the photometry of the rest--frame LyC. In the following section, we
discuss the additional selection criteria aiming at identifying a sample
of isolated SFGs.

\begingroup
\tiny
\begin{longtable}{cccc}
\caption{Archival GALEX FUV Imaging}\\
\hline \hline
\multicolumn{1}{c}{{\it Field}} &
\multicolumn{1}{c}{$N_{\mbox{\it \tiny gal}}$}  &
\multicolumn{1}{c}{$t_{\mbox{\it \tiny eff}}$[ks]}  &
\multicolumn{1}{c}{$\bar{m}_{\mbox{\it \tiny FUV,sky}}$}  \\ \hline \hline
\endhead
\multicolumn{4}{l}{{\sc Notes:}}\\
\multicolumn{4}{l}{Col.\! 2: Number of galaxies, per field}\\
\multicolumn{4}{l}{Col.\! 3: Effective exposure time, in ks,}\\
\multicolumn{4}{l}{\,\,\,\,\,\,\,\,\,\,\,per galaxy}\\
\multicolumn{4}{l}{Col.\! 4: 3$\sigma$ FUV magnitude limit}\\
\multicolumn{4}{l}{\,\,\,\,computed in a aperture with a radius}\\
\multicolumn{4}{l}{\,\,\,\,equal to 0.67$\times$4\farcs5.}\\
\endlastfoot 
AEGIS & 219 & 146 & 27.2  \\ 
COSMOS & 132 & 206 &  27.4  \\ 
GOODS-N & 113 & 126 & 27.1  \\ 
GOODS-S & 198 & 104 & 27.2  \\ 
\hline \hline \label{tab:simppartab}
\end{longtable} 
\endgroup 

First, we identify isolated SFGs applying the same selection criteria
applied by CBT09. Specifically, we remove any galaxy within 16$\farcs$0
of a galaxy or star brighter than FUV=23 or within 8$\farcs$0 of a
23$<$FUV$<$25 galaxy. In both instances, only those galaxies at $z<$0.7
are considered. To measure photometry for sources in the GALEX imaging,
we used \sex~in dual--image mode, with the detection made in the NUV
band.  These criteria select from the initial sample a subset of 1050
($\sim$80\%) of the initial sample of all \rsrang\,ELGs. For reference,
this sample is $\sim\,1.5\times$ larger than that considered by CBT09. 

A visual inspection reveals that galaxies selected by the CBT09 criteria
alone may still be in close proximity to other potentially low--redshift
sources, making the unambiguous measurement of LyC photometry for
the SFGs in the sample difficult (see Figure \ref{fig:gradeb}). Thus, 
we make a third and more stringent selection against SFGs with nearby
neighbors, removing {\it all} SFGs with sources at $z<0.9$ within
3\farcs0. 

Together these criteria define our ``Grade A'' sample, containing 618
SFGs.  The 1050 SFGs selected with the CBT09 criteria will be referred
to as the ``Grade B'' sample in the following analysis.  We provide the
number of Grade A and B galaxies in each field in Table
\ref{tab:simppartab}. The emission line selection we used to identify
the galaxy sample enables us to identify a large number of dwarf
galaxies,   in a mass range  (50\% of the sample is below M$=10^9$\Msol)
that can be as much as $\sim$1.5 dex {\it lower} than the characteristic
stellar mass of $z\sim1$ SFGs (e.g., van der Burg 2013). 

With the large sample size of Grade A SFGs we define three sub--samples,
each including hundreds of galaxies (see Table \ref{tab:longtab1}),
with similar physical characteristics. Comparisons of \fesc~ measured
for galaxies between these sub--samples may allow us to probe in unique
an novel ways the efficiency with which ionizing photons escape from
SFGs.  

\begin{enumerate}

\item {\bf H$\alpha$ Equivalent Width} --- We identify  ``extreme
emission line'' galaxies (EELGs; see Atek et al.~2011), requiring
rest--frame $EW_{\mbox{\tiny H$\alpha$}}\ge200\mbox{\AA}$.  These galaxies populate a unique parameter
space in the study of LyC escape in which strong starbursts necessarily occur in
relatively shallow gravitational potentials. These data will provide the
first constraints on \fesc~for such galaxies at this redshift range.

\item {\bf Orientation} --- We distinguish face-on galaxies from the
full sample of Grade A SFGs using the ellipticity measured in the F140W
band.  We use the major--to--minor axis ratio criteria, $1-\frac{b}{a}<$0.15, 
which is a good proxy for inclination angle (Maller et al. 2009) to
identify edge/face--on SFG. A distinction between \fesc~measured between
these two classes can provide a direct test whether LyC photons escape
isotropically from the disk or instead preferentially through channels
in the ISM.

\item {\bf Stellar Mass or Luminosity} --- We select equally--sized
samples of low, medium and high ($\bar{M}\simeq$ 8.9, 9.5 and 10.5
\Msol, respectively) mass SFGs, using stellar masses reported in the
3DHST catalogs.  Similarly, we define three bins in luminosity with the
median absolute magnitude of each bin equal to $-17.1,-18.1,$ and $-18.9$.
Dahlen et al.\,(2007) report a characteristic UV magnitude for the
$z\sim1$ UV luminosity function equal to $M^*=19.98$ mag; thus the
median UV magnitudes of these three bins are equivalent to 0.07, 0.18,
and 0.37$\times\!L^*$.

\end{enumerate} 

The correlation between these characteristics and the detection (or lack
of) of escaping LyC emission enhances our understanding of the
physics of LyC escape. Here, we qualify each of these characteristics
and their use in better revealing those factors affecting the escape of
LyC from SFGs.

First, EELGs are defined on the basis of their high emission-line
equivalent widths, indicative of a recent episode of star formation.
EELGs are characterized by high specific star-formation rates
(\gsim$10^{-8}$yr$^{-1}$), and may be offset by as much as $\sim1$ dex
from the ``star-forming'' main-sequence of galaxies at $z\sim1$ (Atek et
al.~2011; Wuyts et al.~2011; Amor\'{i}n et al.~2014).   At high
redshift, EELGs have been observed to likely emit LyC emitters (see
recent work by de Barros 2015) and thus are of interest to consider
in this work specifically.  Specifically, extreme H$\alpha$-ELGs are of interest, which can be detected in G141 at $z\sim1$. At
$z\sim0.2$, Cowie et al. (2011) found $\sim$75\% of Ly$\alpha$ emitters
selected from GALEX grism observations to have EW(H$\alpha$)$>$100\AA.
More recently, Henry et al. (2015) observed 10 high EW, luminous compact
SFGs with UV HST Cosmic Origins Spectrograph (COS) observations and
found Ly$\alpha$ emission in 100\% of the galaxies.  Noting that neutral
hydrogen is optically thick to Lyman-$\alpha$ at column densities 10$^4$
times {\it lower} than at the Lyman-limit, these observed correlations
between Extreme-H$\alpha$ ELGs and Ly$\alpha$ make EELGs appealing
candidates in a search for escaping LyC emission.

The second criterion ---galaxy inclination--- seeks to characterize
the role of anisotropic escape. Assuming that the LyC--emitting
star--forming regions (cf., Kimm and Cen 2014) are sequestered to a
gas--rich disk in the galaxy, the column of neutral gas and dust in
the ISM is lowest in the direction normal to the disk.  Channels through the 
attenuating medium are evacuated more easily perpendicularly to the disk. As a result,  \fesc~could
show a dependence on the orientation of the disk relative to the
observer.  


The final selection criterion is designed to constrain \fesc~as a
function of UV luminosity, because it has been suggested that LyC may
easily escape from low mass SFGs. In fact, the CANDELS survey
demonstrates that the observed populations of SFGs at $z>6$ do not,
under reasonable assumptions, produce sufficient ionizing photons to
maintain an ionized universe (e.g., Robertson et al. 2015, Finkelstein
et al. 2015). To resolve this tension it has been suggested that
\fesc~is higher for low luminosity galaxies, and a population of very
faint ($M_{UV}$\lsim$-13$), yet unobserved galaxies with high
\fesc\gsim20\% is typically invoked. Our sample probes two magnitudes
fainter in the UV than previous studies at similar redshifts, thus
allowing us to investigate any dependency of \fesc~with $M_{UV}$. We
refer the reader to Figure 2 for an illustration of the stellar masses
to which we are sensitive with these data. 

\begin{figure}[h!] \begin{center}
\includegraphics[scale=0.45]{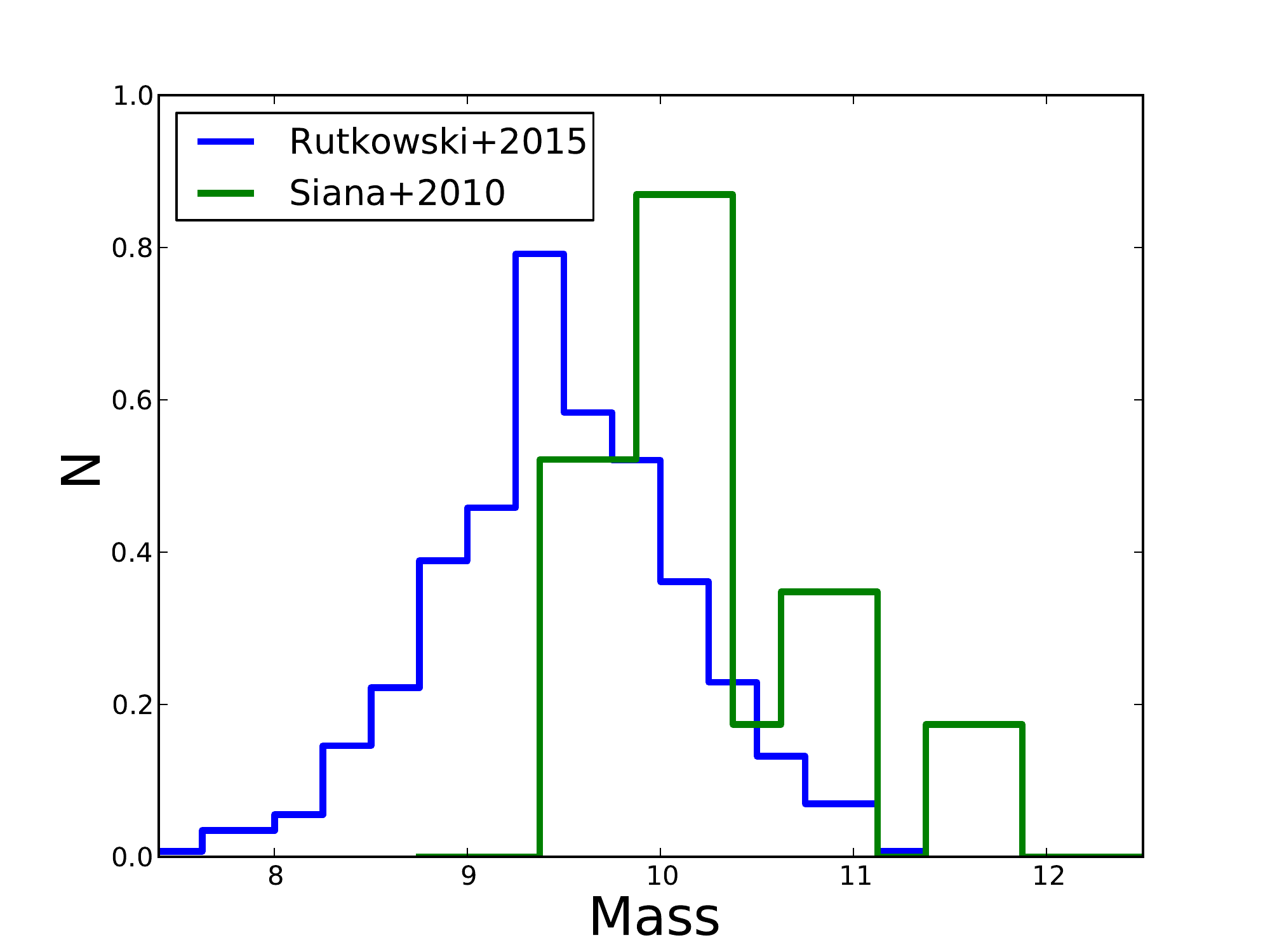}

 \caption{The normalized distribution of stellar masses for the Grade A
 sample (blue). Illustrated for comparison are the normalized
 distributions of stellar mass for the Siana et al. (2010) $z\sim1$
 sample of SFGs (green). Our selection on emission--line galaxies, in
 contrast to, e.g., the strength of the non--ionizing UV continuum, is
 largely insensitive to the stellar mass of the galaxy. Thus, we are
 able to extend the search for escaping LyC from $z\sim1$ SFGs to
 significantly lower stellar masses than previously considered. For
 example, the median mass of the Grade A SFGs equals $10^{9.3}$[\Msol];
 $\sim50\%$ of these galaxies are more than an order of magnitude {\it
 less massive} than considered in previous work.}
  \label{fig:massageAB}
 \end{center}
\end{figure}


%
%
%
%

\section{Observed UV to LyC flux ratio}

For each galaxy, we use ground--based $U$-band and GALEX FUV photometry
to compute the observed non-ionizing (UV) to  LyC flux ratio,
$(L_{UV}/L_{LyC})_{obs}$. The 3$\sigma$ magnitude limit of the LyC
images (GALEX FUV)  ranges between 27.4 and 27.1 (see Table~1). Thus,
considering the average UV -- LyC colors of known LyC leakers (see e.g.,
Nestor et al. 2013, Mostardi et al. 2015), our survey is sensitive to
LyC radiation in individual objects with rest-frame UV magnitude (i.e.,
$U-$band magnitude) brighter than 25.5.  Thus, individual candidates
may be detected in the GALEX FUV images sensitive to rest-frame LyC. 

We first visually inspected all of the ELGs' GALEX FUV stamps to search
for these potential individual detections, and find 6 ELGs selected
using the CBT09 criteria (i.e., Grade B ELGs) are identified.  We used
Source Extractor in dual image mode---the corresponding stacked GALEX
NUV images were used for the definition of the sources' Kron
apertures---to confirm that each of these sources were detected at
\gsim\,$2.5\sigma$ in the LyC images. In each case, a comparison of
these galaxies with the higher resolution HST F140W and ground-based
$U$-band data\footnote{We use the native scale $U$-band images
downloaded from the CADC archives at
http://www4.cadc-ccda.hia-iha.nrc-cnrc.gc.ca/en/cfht/}, strongly
suggests that the flux is not associated with the SFG, but rather with a
foreground galaxy in close proximity.  No galaxies in the more robust
Grade A sample are detected individually in the LyC image. 
 
\begin{figure}[htb] \centering
\subfigure{\label{fig:a1}}\includegraphics[scale=0.45]{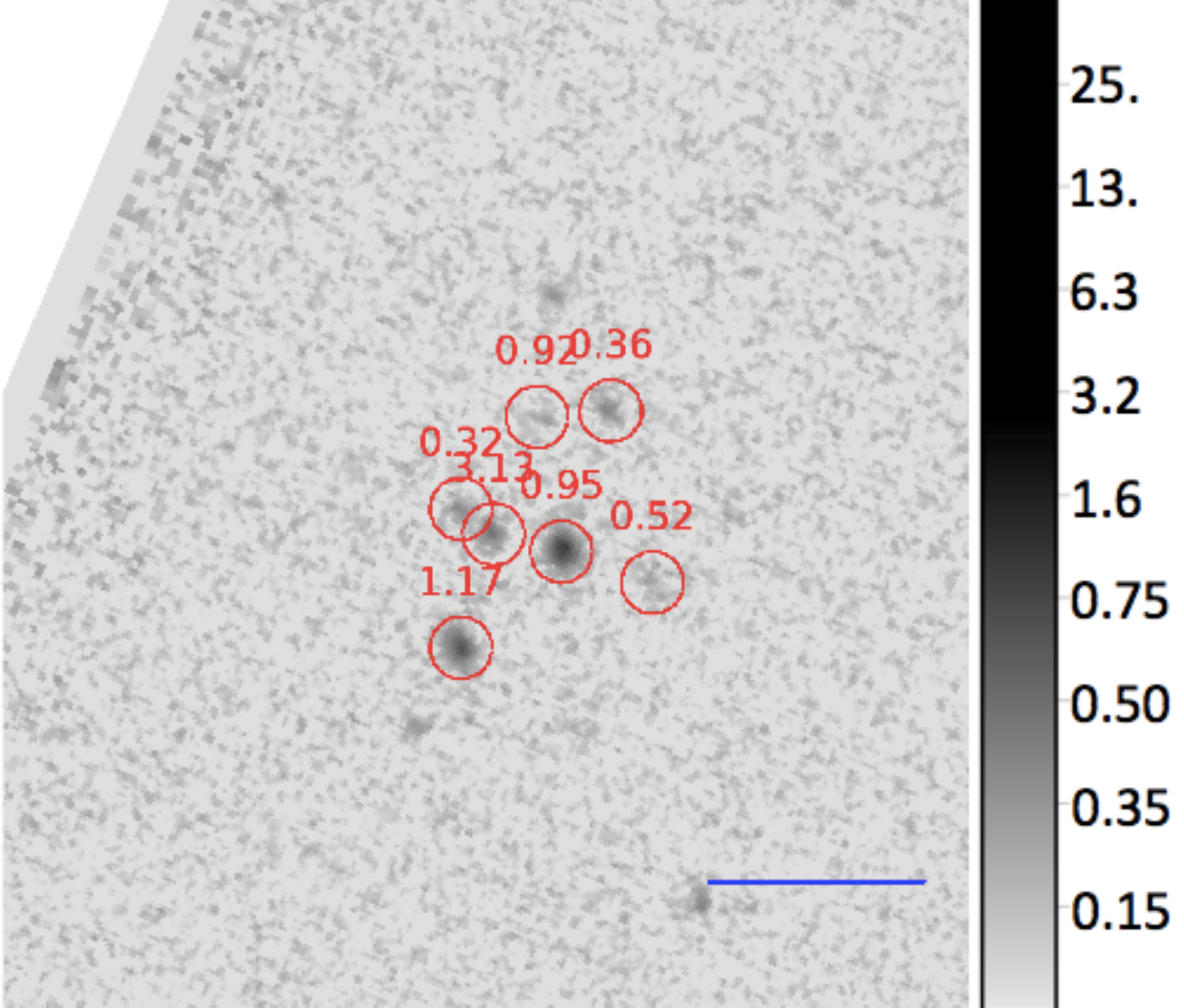}\\
\subfigure{\label{fig:a2}}\includegraphics[scale=0.45]{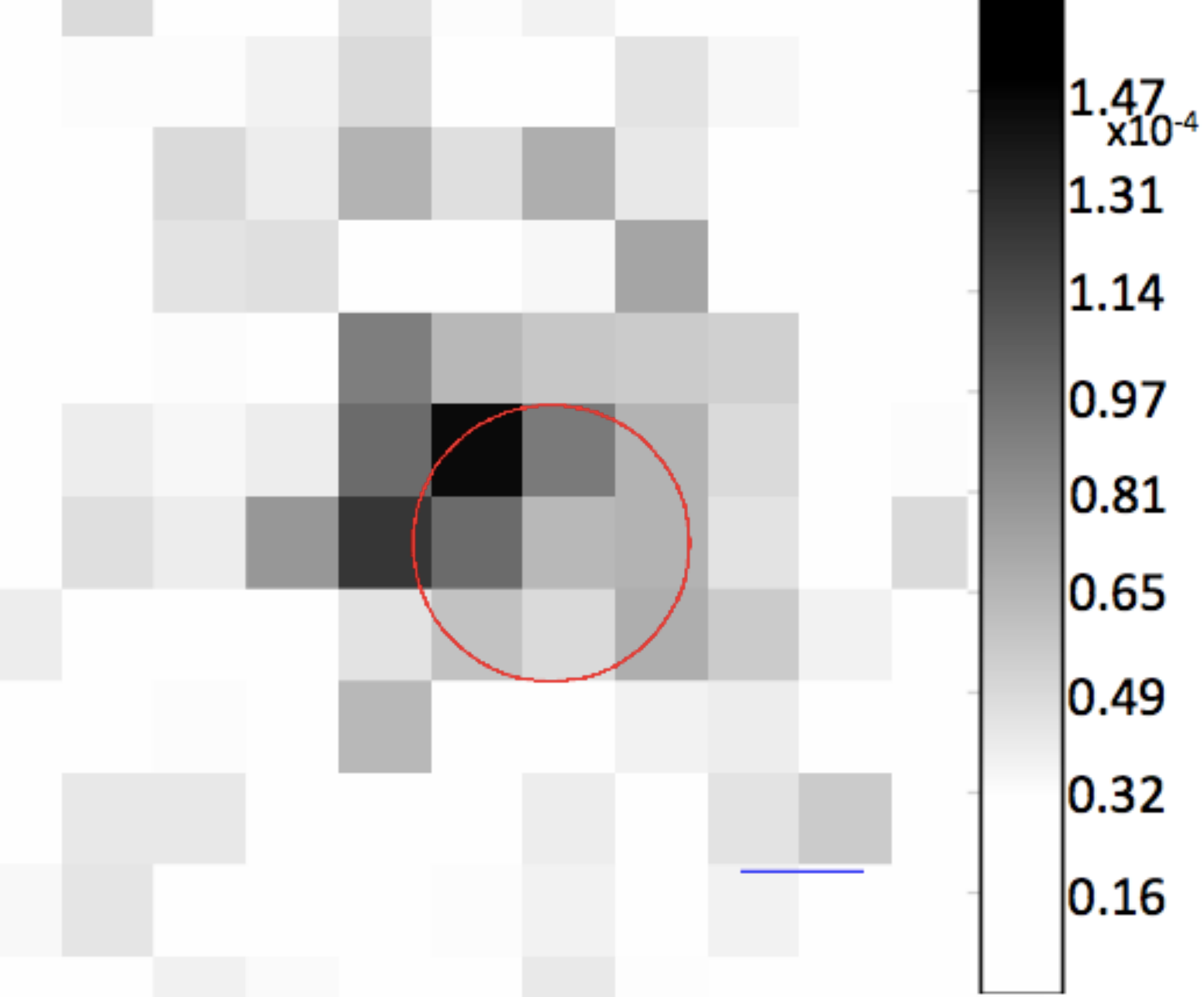}\\
\caption{Approximately 30\% of the ELGs identified by the CBT09
selection criteria are in close spatial proximity to low-redshift
($z<0.9$) galaxies which may contaminate the UV photometry. Here we
illustrate an galaxy in the COSMOS field, J100021.96+021545.6. In
the upper panel is shown a $9\farcs\times9\farcs$ logarithmically-scaled HST F140W postage stamp, centered on the $z=0.95$ SFG. Overplotted are red
regions ($r$=0$\farcs$5) identifying sources identified in close
proximity, with the labels indicating the galaxy's photometric redshift.
Similarly in the lower panel, the corresponding GALEX FUV postage stamp
is illustrated. Here, a red region (r=$2\farcs25$; 0.5$\times$ the GALEX
FUV $4\farcs5$ FWHM) is centered on the position of the galaxy.  In both
panels the blue scale bar has a fixed length of 2$\farcs$0 and units are cts s$^{-1}$}
\label{fig:gradeb} \vspace{+10pt} \end{figure}

\begin{figure}[htbc] \begin{center}
\includegraphics[scale=0.35]{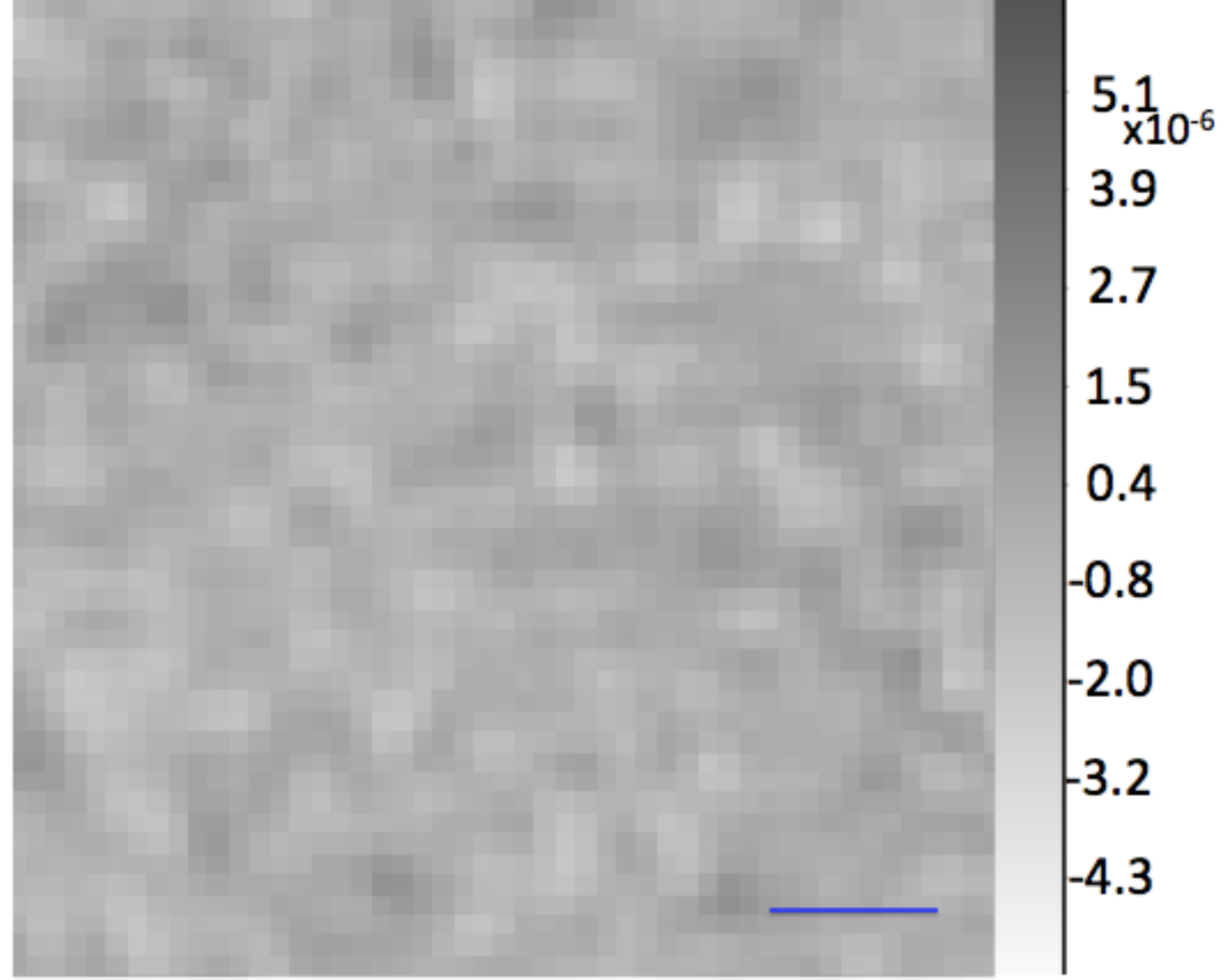} \caption{The
Grade A stacked GALEX FUV image of ELGs at $z\sim1$, linearly scaled
(cts $s^{-1}$) and smoothed with a $\sigma=2pix$ Gaussian kernel, shows
no significant (3$\sigma$) emission.  Note, the scalebar here has a
length of 15\farcs. We produce stacked images for all subsamples of ELGs
listed in Table 2 and measure the variance in the sky background in
1000s of randomly-placed circular apertures to determine upper limits to
the LyC flux.  The upper limits to relative escape fraction of LyC
presented in Table 3 were the derived from these measurements.}
\label{fig:stacked} \end{center} \end{figure}

With no unambiguous individual object detected in LyC, we perform a
stacking analysis, averaging the FUV images of all  galaxies, as well as
of the objects in the subsamples described in Section~4. When creating
the stack for the Grade~B sample, we removed those objects that had a
spurious individual detection in the FUV. To prepare the stacks, we
first produced equally sized postage FUV images centered on the SFG and
then median combined all stamps, scaling the flux in each stamp by the
position-dependent exposure time. For each subsample, we measured the
total flux within an optimally-sized aperture of radius
$0.67\times$4\farcs2. In Table~3 we report the 3$\sigma$ flux limits of
the stacked FUV LyC images for each subsample. We find no significant
detection, above 3$\sigma$, in the stacked images (see Figure
\ref{fig:stacked}). 
 
At the redshift of our sample, the non--ionizing rest-frame 1500\AA\
continuum is covered by the $U$-band filter, which is minimally affected
by Ly$\alpha$ emission and Ly$\alpha$--forest absorption. For each
subsample, we computed the median $U-$band flux density of all objects
in the sample, and report  the measurements in Table~3. All measurements
in Table \ref{tab:longtab1} are corrected for (minor) Milky Way
extinction at non-ionizing and ionizing UV continuum wavelengths, using
extinction laws defined by \cite{CCM89} and \cite{Odonnell94}.  We
correct for Galactic extinction assuming a spatially uniform color
excess specific to each field from the dust maps of Schlegel, Finkbeiner
and Davis (1998). In Figure 4, we compare the $3\sigma$
limit on the observed LyC-to-UV flux ratio, corrected for average
attenuation along the line of sight (($L_{\rm UV}/L_{\rm
LyC})_{obs}\exp[\tau_{-HI,IGM}]$), computed for our Grade~A and high--EW
samples. For reference, we include the observed IGM corrected ratios
from previous surveys of {\it individual} $z\sim1$ galaxies.  The large
number of galaxies that enter in the calculation of the 3$\sigma$ upper
limits allow us to probe a factor of a few deeper in )$L_{\rm UV}/L_{\rm LyC}$)$_{obs}$  than previous studies.

\begin{figure}[h!] \begin{center}
\includegraphics[scale=0.42,trim=+1cm 0 0 0]{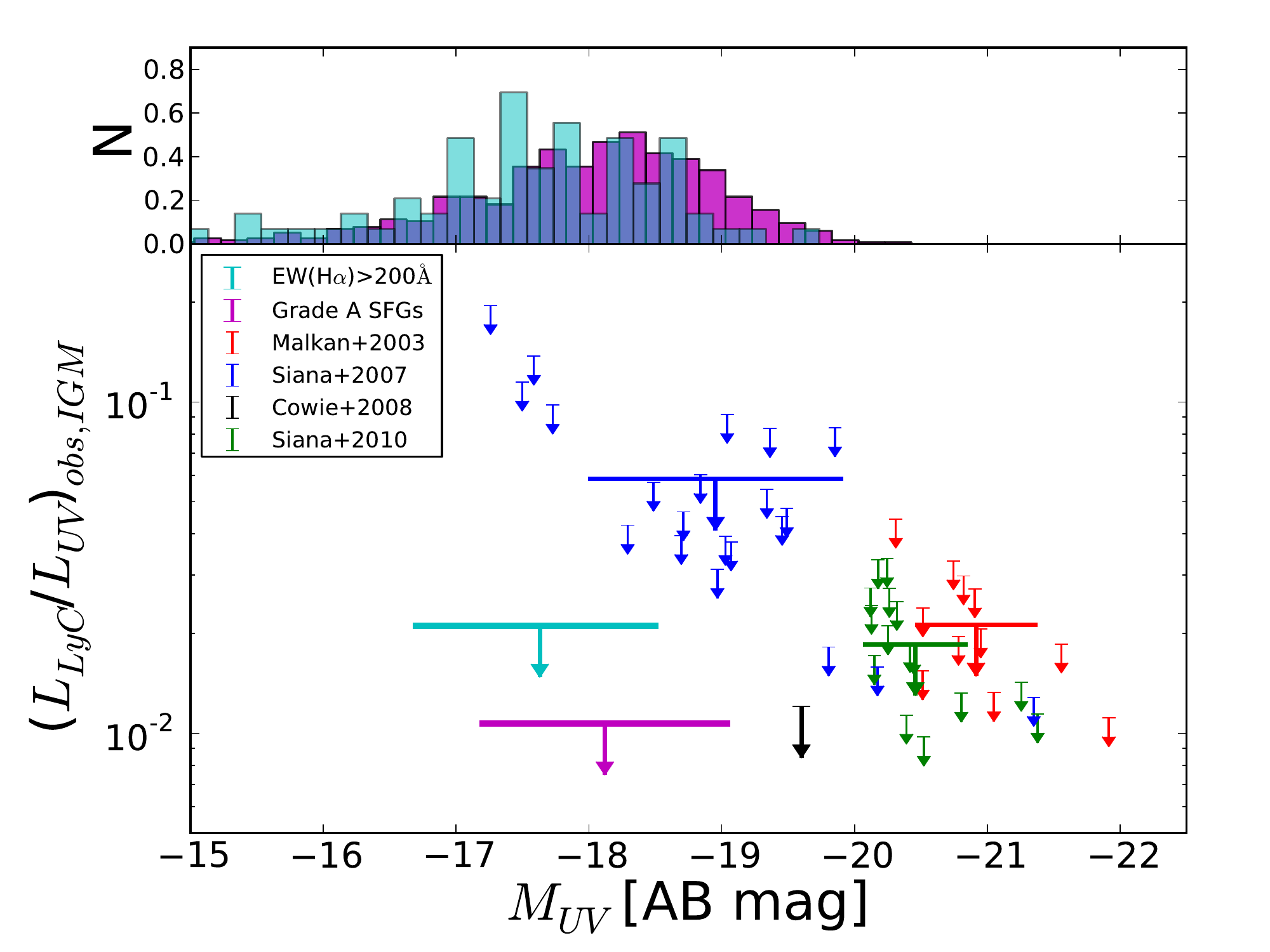} \caption{The observed LyC-to-UV flux
ratio, corrected for IGM extinction, is provided here for the Grade A
and High EW samples (violet and cyan, respectively). The range on the
absolute magnitude provided indicates the 1$\sigma$ width on the median
for each sample, assuming the sample distribution is Gaussian.  For
reference, we also reproduce the upper limits measured for individual
$z\sim1$ SFGs from Malkan et al. (2003), and Siana et al. (2007,2010) in
red, green, and blue limits, respectively. Stacking GALEX FUV data
enables us to place limits to the UV continuum ratio for $z\sim1$ SFGs
as good or better as in previous studies, and we can extend previous efforts (black, CBT09) to  
significantly fainter populations of SFGs.} \end{center} \end{figure}

\section{The LyC escape fraction}

Lyman continuum emission produced by young, hot stars  embedded in the
gas-rich media of galaxies is attenuated by neutral hydrogen (both
within and outside galaxies) and the dust associated with the
interstellar medium. The {\it absolute} fraction of LyC radiation
escaping from a galaxy is defined as $f_{esc}  \equiv
\frac{L_{LyC,esc}}{L_{LyC,int}} $, where $L_{LyC,esc}$ is the ionizing
luminosity attenuated by the gas and dust in the interstellar medium, and
$L_{LyC,int}$ is the intrinsic emissivity of the stars below $912$\AA.
At redshifts $z\gtrsim 0.4$, the contribution by the neutral
intergalactic medium cannot be neglected and the observed luminosity below 912\AA\ can be written as
$L_{LyC,obs}=L_{LyC,esc}\exp{[-\tau_{HI,IGM}]}$. With these definitions,
the absolute escape fraction of ionizing radiation can be written in
terms of the observed ionizing emissivity as:\vspace{-10pt}

\begin{center}
\begin{equation}\label{eqn:eq1} 
f_{esc} = \frac{L_{LyC,obs}\exp{[\tau_{HI,IGM}]}}{L_{LyC,int}}.
\end{equation}
\end{center}\vspace{+5pt}
If the attenuation by the galaxy's neutral hydrogen and dust ($\exp[{-\tau_{HI,ISM}}]$ and $\exp[{-\tau_{dust,\lambda}}]$, respectively) were known, the absolute escape fraction could  be written as:\vspace{-10pt}
\begin{center}
\begin{equation}\label{eqn:eq2} 
f_{esc} = \exp[{-\tau_{HI,ISM}}]\cdot\exp[{-\tau_{dust,\lambda}}].
\end{equation}
\end{center}\vspace{+5pt}
However, the dust extinction curve at wavelengths lower than 912\AA\  ($\tau_{dust,\lambda}$) is highly uncertain. 
Moreover, the attenuation of $L_{LyC,int}$  by neutral gas and dust (i.e., $\exp[{-\tau_{HI,ISM}}]$ and 
$\exp[{-\tau_{dust,\lambda}}]$) within the  SFG population varies strongly with the total gas and dust mass, the
``patchiness" and distribution of gas and dust, and the orientation of the disk with respect to the observer.  
To overcome these uncertainties, Steidel et al. (2001) introduced the {\it relative} escape fraction, defined as:\vspace{-10pt}
\begin{center}
\begin{equation} 
f_{esc,rel}=f^{LyC}_{esc,rel}=\frac{(L_{UV}/L_{LyC})_{int}}{(L_{UV}/L_{LyC})_{obs}}\cdot\exp[{\tau_{IGM}}],
\end{equation}\label{eqn:eq3}
\end{center}\vspace{-10pt}

\noindent corresponding to the fraction of emitted 900\AA\ photons that
escapes the galaxy without being absorbed by interstellar medium divided
by the fraction of non-ionizing (at 1500\AA) photons that escapes. Note
that the relative escape fraction in Equation 3 is corrected for HI
attenuation by the IGM.  If the column density and redshift
distributions of neutral absorbers were known, the opacity of the IGM
along the line of sight could be modeled. However, because this is
typically not the case, a statistical redshift--dependent correction is
usually applied to derive $L_{LyC,esc}$ from $L_{LyC,obs}$ (Madau et al.
1996, Haardt and Madau et al. 2012).

The relative escape fraction can be directly related to \fesc$:$\vspace{-10pt}

\begin{center}
\begin{equation}  
f_{esc} = f_{esc,rel}\cdot\exp[{-\tau_{dust,UV}}],
\end{equation} \label{eqn:eq4}
\end{center}\vspace{-10pt}
\noindent where  $\exp[{-\tau_{dust,UV}}]$ is the attenuation of the non-ionizing
UV continuum (typically measured at $1500\mbox{\AA}$) by dust.  

\begin{figure*}[htb]
\begin{longtable}{lccccccc} 
\caption{Upper limits to \fesc~at $z\sim$1, Measured Quantities}\\
\hline \hline \\[-2ex]
\multicolumn{1}{l}{{\it Selection}} &
\multicolumn{1}{l}{N$_{gal}$} &
\multicolumn{1}{c}{$\bar{f}_U$ ($\bar{m}_{U}$)$^a$ } & 
\multicolumn{1}{c}{$\bar{f}_{LyC}$ ($\bar{m}_{LyC}$)$^a$}  & 
\multicolumn{1}{c}{$\left(\frac{L_{UV}}{L_{LyC}}\right)_{obs}$} &
\multicolumn{1}{c}{$\left(\frac{L_{UV}}{L_{LyC}}\right)_{IGM,corr}$} &
\multicolumn{1}{c}{$\left(\frac{L_{LyC}}{L_{UV}}\right)_{obs}$} &
\multicolumn{1}{c}{$\left(\frac{L_{LyC}}{L_{UV}}\right)_{IGM,corr}$} \\ 
\multicolumn{1}{l}{} &
\multicolumn{1}{l}{} &
\multicolumn{1}{c}{[Jy (mag)]} & 
\multicolumn{1}{c}{[Jy (mag)]} & 
\multicolumn{1}{c}{} &
\multicolumn{1}{c}{} &
\multicolumn{1}{c}{} &
\multicolumn{1}{c}{} \\ [0.5ex]\hline \hline \\[-1.8ex]
\endhead
\multicolumn{8}{l}{{\sc Notes:} Here we accumulate upper limits (3$\sigma$) which are directly measured from the UV photometry.  Columns entries}\\
\multicolumn{8}{l}{\,\,\,are defined as follows$:$}\\
\multicolumn{8}{l}{Column 1: Sample Selection criteria (see Section 4);}\\
\multicolumn{8}{l}{Column 2: Number of galaxies included in each galaxy. The mass and luminosity sub-samples were selected to have equal}\\
\multicolumn{8}{l}{\,\,\,\,\,\,numbers of galaxy per bin;}\\
\multicolumn{8}{l}{Columns 3,5: The observed UV continuum ratios;}\\
\multicolumn{8}{l}{Columns 4,6: The UV continuum ratios corrected for a mean transmission of the IGM of 55\% at $z\sim1$ (Section 7).}\\
\multicolumn{8}{l}{\,\,\,We note that although the Grade A sample includes nearly the same number of SFGs (618 vs. 626) considered in CBT09,}\\
\multicolumn{8}{l}{\,\,\,the upper limits presented here for the LyC luminosity are $\sim$50\% deeper due to the availability of $\sim$5 years of well-calibrated}\\
\multicolumn{8}{l}{\,\,\,deeper, co-added images than were available in the MAST archive for that novel study of the}\\
\multicolumn{8}{l}{\,\,\,escape of LyC at $z\sim1$ in UV archival images.}\\
\endlastfoot
Grade A & 618 & 6.5$\times10^{-7}$ (24.43) & 2.7$\times10^{-9}$ (30.42) & 235.1 & 129.3 & 4.2$\times10^{-3}$ & 0.7$\times10^{-2}$\\
Grade B & 997 & 6.5$\times10^{-7}$ (24.44) & 2.1$\times10^{-9}$ (30.68) & 295.8 & 162.7 & 3.3$\times10^{-3}$ & 0.6$\times10^{-2}$\\
Edge-On & 455 & 6.3$\times10^{-7}$ (24.47) & 3.2$\times10^{-9}$ (30.26) & 195.8 & 107.6 & 5.1$\times10^{-3}$ & 0.9$\times10^{-2}$\\
Face-On & 118 & 7.4$\times10^{-7}$ (24.29) & 5.9$\times10^{-9}$ (29.60) & 125.1 & 68.83 & 7.9$\times10^{-3}$ & 1.4$\times10^{-2}$\\
High Mass & 206 & 7.9$\times10^{-7}$ (24.22) & 4.8$\times10^{-9}$ (29.82) & 164.9& 90.72 & 6.0$\times10^{-3}$ & 1.1$\times10^{-2}$\\
Med. Mass & \dittotikz & 7.3$\times10^{-7}$ (24.31) & 4.7$\times10^{-9}$ (29.84) & 154.5 & 85.01 & 6.4$\times10^{-3}$ & 1.1$\times10^{-2}$\\
Low Mass & \dittotikz & 4.3$\times10^{-7}$ (24.87) & 4.6$\times10^{-9}$ (29.86) & 93.83 & 51.61 & 10.6$\times10^{-3}$ & 1.9$\times10^{-2}$\\
High EW & 72 & 6.5$\times10^{-7}$ (24.43) & 7.4$\times10^{-9}$ (29.35) & 88.26& 48.54 & 11.3$\times10^{-3}$ & 2.0$\times10^{-2}$\\
Bright M$_{UV}$ & 206 & 1.1$\times10^{-7}$ (23.80) & 4.7$\times10^{-9}$ (29.83) & 245.4 & 135.0 & 4.0$\times10^{-3}$ & 0.7$\times10^{-2}$\\
Med. M$_{UV}$ & \dittotikz & 5.4$\times10^{-7}$ (24.63) & 4.7$\times10^{-9}$ (29.85) & 115.1 & 63.31 & 8.6 $\times10^{-3}$ & 1.5$\times10^{-2}$\\
Faint M$_{UV}$ & \dittotikz & 2.4$\times10^{-7}$ (25.50) & 4.6$\times10^{-9}$ (29.87) & 53.10 & 29.20 & 18.8$\times10^{-3}$ & 3.4$\times10^{-2}$\\ \hline \hline
\label{tab:longtab1}
\end{longtable}
\end{figure*}

\section{Observational Constraints on the escape fraction}\label{subsec:fescp1}
\subsection{Absolute \fesc~from relative \fesc}
\label{sec:fescrel}

We use Equations~3 and 4 to compute upper limits on the relative and
absolute escape fractions. The intrinsic $(L_{UV}/L_{LyC})_{int}$ ratio
is highly uncertain and depends on various properties of the stellar
population (see discussion in Section~\ref{sec:intratio}). We assume an
average value of $(L_{UV}/L_{LyC})_{int}=7$ (in L$_{\nu}$). This value
was computed from constant star-formation Starbust99 models, assuming a
Salpeter IMF, solar metallicity and no stellar rotation. This ratio is
consistent with what is typically assumed in the literature; e.g., Siana
et al. (2010) use $(L_{UV}/L_{LyC})_{int}\simeq$8.

We compute the correction factor for absorption by the IGM
($\exp[\tau_{HI,IGM}]$), from the piecewise parametrization of the
distribution in redshifts and column density of intergalactic absorbers
as described in Haardt and Madau (2012). We derive a mean IGM
transmission at $\lambda<912\mbox{\AA}$ of T$\simeq$55\% (corresponding
to $\exp[{\tau_{IGM}}]\sim 1.8$), computed at the median redshift of our
sample. We verified that the assumed value of the IGM opacity does not
introduce systematics in our calculations: the redshift distribution is
approximately flat in the 0.9--1.4 redshift range, and the attenuation
changes by less than 5\% at the two extrema of the redshift range. This
small variation is the result of the shallow redshift dependency of the
column density distribution of absorbers, which scales as $(1+z)^{0.16}$
in the redshift range $0<z<1.5$ (Weymann et al. 1998).

We report the relative escape fraction calculated assuming
$(L_{UV}/L_{LyC})_{int}=7$ in Table~\ref{tab:longtab1}; the range of
possible values to this intrinsic ratio is discussed further in the
Subsection 7.1.1. In order to compute the absolute escape fraction
($f_{esc}$) we need to correct the relative escape fraction for dust
attenuation at 1500\AA. We computed an average attenuation for each
group, using the stellar $E(B-V)_s$ provided by Skelton et al. (2014),
assuming a Calzetti extinction law (Calzetti et al. 2000). The absolute
escape fractions are also reported in Table~\ref{tab:longtab1}.

\subsubsection{The intrinsic ($L_{UV}/L_{LyC}$)$_{int}$ ratio}
\label{sec:intratio} The values of the absolute escape fraction reported
in column 3 of Table~\ref{tab:longtab2} depend on two assumptions:
the average IGM attenuation (discussed in the previous section) and the
intrinsic UV--to--LyC ratio.  Here, we discuss the
systematic uncertainty in this intrinsic ratio arising from the choice
of stellar models.

We have assumed  ($L_{UV}/L_{LyC}$)$_{int}=7$, consistent with previous
studies in the literature. For a given star-formation history, this
ratio depends strongly on the age of the stellar population, the initial
mass function, as well as whether or not rotation accounted for in the
stellar libraries used to compute the synthetic spectra (Levesque et
al. 2012, Leitherer et al. 2014, Dominguez et al. 2015).  We have
measured the systematic uncertainty associated with the assumption of a constant
value, using a large library of  galaxy spectra generated with two
stellar population synthesis codes, {\it galaxev} (Bruzual and Charlot
2003) and {\it Starburst99} \citep{Leitherer99}\footnotetext{Starburst99
is available online at$:$
http://www.stsci.edu/\\science/starburst99/docs/default.htm.} that
incorporate a variety of stellar templates which have varied intrinsic
LyC-to-UV continuum ratios as we will demonstrate below.

\begin{figure*}[htb]
\begin{center}
\begin{longtable}{lccccccc} 
\caption{Upper limits to \fesc~at $z\sim$1, Derived Quantities}\\
\hline \hline \\[-2ex]
\multicolumn{1}{l}{{\it Selection}} &
\multicolumn{1}{c}{$f_{esc,rel}$} & 
\multicolumn{1}{c}{$f_{esc}$} &
\multicolumn{1}{c}{$L_{H\alpha}$} &
\multicolumn{1}{c}{$Q_{ion}$} &
\multicolumn{1}{c}{$\nu_{\mbox{eff}}L_{LyC,int}$} &
\multicolumn{1}{c}{$\nu\,L_{LyC,obs}$} &
\multicolumn{1}{c}{$f_{esc}^{H\alpha}$} \\
\multicolumn{1}{l}{} &
\multicolumn{1}{c}{} & 
\multicolumn{1}{c}{} &
\multicolumn{1}{c}{[erg s$^{-1}$]} &
\multicolumn{1}{c}{[photons s$^{-1}$]} &
\multicolumn{1}{c}{[erg s$^{-1}$} &
\multicolumn{1}{c}{[erg s$^{-1}$]} &
\multicolumn{1}{c}{} \\ [0.5ex]\hline \hline \\[-1.8ex]
\endhead
\multicolumn{8}{l}{{\sc Notes:} The upper limits to \fesc~derived from the ionizing to non--ionizing UV continuum}\\
\multicolumn{8}{l}{\,\,\,with columns defined as follows:}\\
\multicolumn{8}{l}{Column 1: Sample Selection criteria (see Section 4) as in Table 2;}\\
\multicolumn{8}{l}{Columns 2,3: Upper limits to the escape fraction derived from the ratio of ionizing }\\
\multicolumn{8}{l}{\,\,\,\,\,\,to non-ionizing continuum ratios.  These data were measured assuming an intrinsic continuum}\\
\multicolumn{8}{l}{\,\,\,\,\,\,ratio of 7. Please see Section 7.1.1 for a discussion of the range of plausible intrinsic ratios}\\
\multicolumn{8}{l}{\,\,\,\,\,\,for young, star--forming galaxies;}\\
\multicolumn{8}{l}{Columns 4,5: The intrinsic H$\alpha$ luminosity and inferred rate of ionizing photons. As in Cols 2,3}\\
\multicolumn{8}{l}{\,\,\,\,\,\,all quantities have been corrected for nebular dust extinction using the median stellar E(B-V)}\\
\multicolumn{8}{l}{\,\,\,\,\,\, measured by Skelton et al. (2014), as outlined in Section 7.1.2.;}\\
\multicolumn{8}{l}{Column 6: The intrinsic LyC as derived from the H$\alpha$ luminosity;}\\
\multicolumn{8}{l}{Column 7: The observed, dust and IGM corrected LyC luminosity measured from the GALEX}\\
\multicolumn{8}{l}{\,\,\,\,\,\,FUV stacked images;}\\
\multicolumn{8}{l}{Column 8: The upper limit to escaped derived by the new method presented in Section 7.2 from}\\
\multicolumn{8}{l}{\,\,\,\,\,\,the LyC-H$\alpha$ ratio}\\
\endlastfoot
Grade A & 8.0\% & 2.1\% & 8.9$\times10^{41}$ & 1.3$\times10^{54}$ & 3.3$\times10^{43}$ & 1.0$\times10^{42}$ & 3.2\% \\
Grade B & 7.5\% & 1.9\% & 8.7$\times10^{41}$ & 1.2$\times10^{54}$ & 3.3$\times10^{43}$ & 9.8$\times10^{41}$ & 2.9\% \\
Edge-On & 8.5\% & 2.2\% & 8.6$\times10^{41}$ & 1.2$\times10^{54}$ & 3.2$\times10^{43}$ & 1.1$\times10^{42}$ & 3.3\% \\
Face-On & 12\.0\% & 2.6\% & 8.9$\times10^{41}$ & 1.3$\times10^{54}$ & 3.3$\times10^{43}$ & 1.8$\times10^{42}$ & 5.3\% \\
High Mass & 9.0\% & 0.7\% & 1.6$\times10^{42}$ & 2.3$\times10^{54}$ & 6.1$\times10^{43}$ & 1.5$\times10^{42}$ & 2.4\% \\
Med. Mass & 10.0\% & 3.2\% & 6.9$\times10^{41}$ & 1.0$\times10^{54}$ & 2.6$\times10^{43}$ & 1.4$\times10^{42}$ & 5.5\% \\
Low Mass & 15.8\% & 10.9\% & 4.3$\times10^{41}$ & 6.3$\times10^{53}$ & 1.6$\times10^{43}$ & 1.4$\times10^{42}$ & 9.0\% \\
High EW & 15.3\% & 9.6\% & 5.9$\times10^{41}$ & 8.7$\times10^{53}$ & 2.2$\times10^{43}$ & 2.1$\times10^{42}$ & 9.3\% \\
Bright M$_{UV}$ & 5.8\% & 1.7\% & 1.4$\times10^{42}$ & 2.0$\times10^{54}$ & 5.4$\times10^{43}$ & 1.5$\times10^{42}$ & 2.7\% \\
Med. M$_{UV}$ & 12.8\% & 3.9\% & 7.2$\times10^{41}$ & 1.0$\times10^{54}$ & 2.7$\times10^{43}$ & 1.4$\times10^{42}$ & 5.2\% \\
Faint M$_{UV}$ & 27.5\% & 5.6\% & 5.9$\times10^{41}$ & 8.7$\times10^{53}$ & 2.2$\times10^{43}$ & 1.4$\times10^{42}$ & 6.1\% \\ \hline \hline 
\label{tab:longtab2}
\end{longtable}
\end{center}
\end{figure*}

\begin{figure}[htbc] 
\begin{center} 
  \includegraphics[scale=0.37]{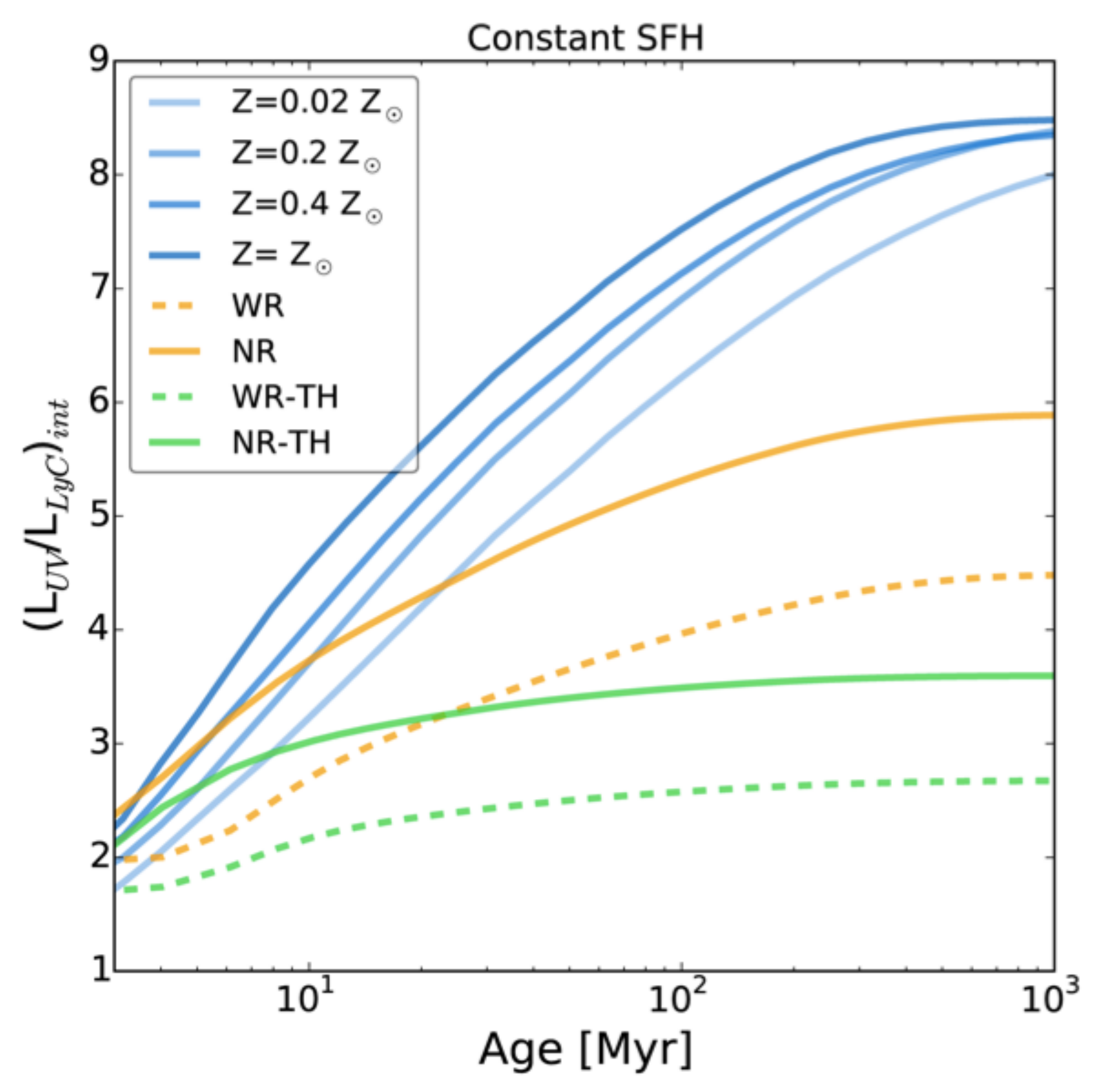} \\
  \includegraphics[scale=0.35]{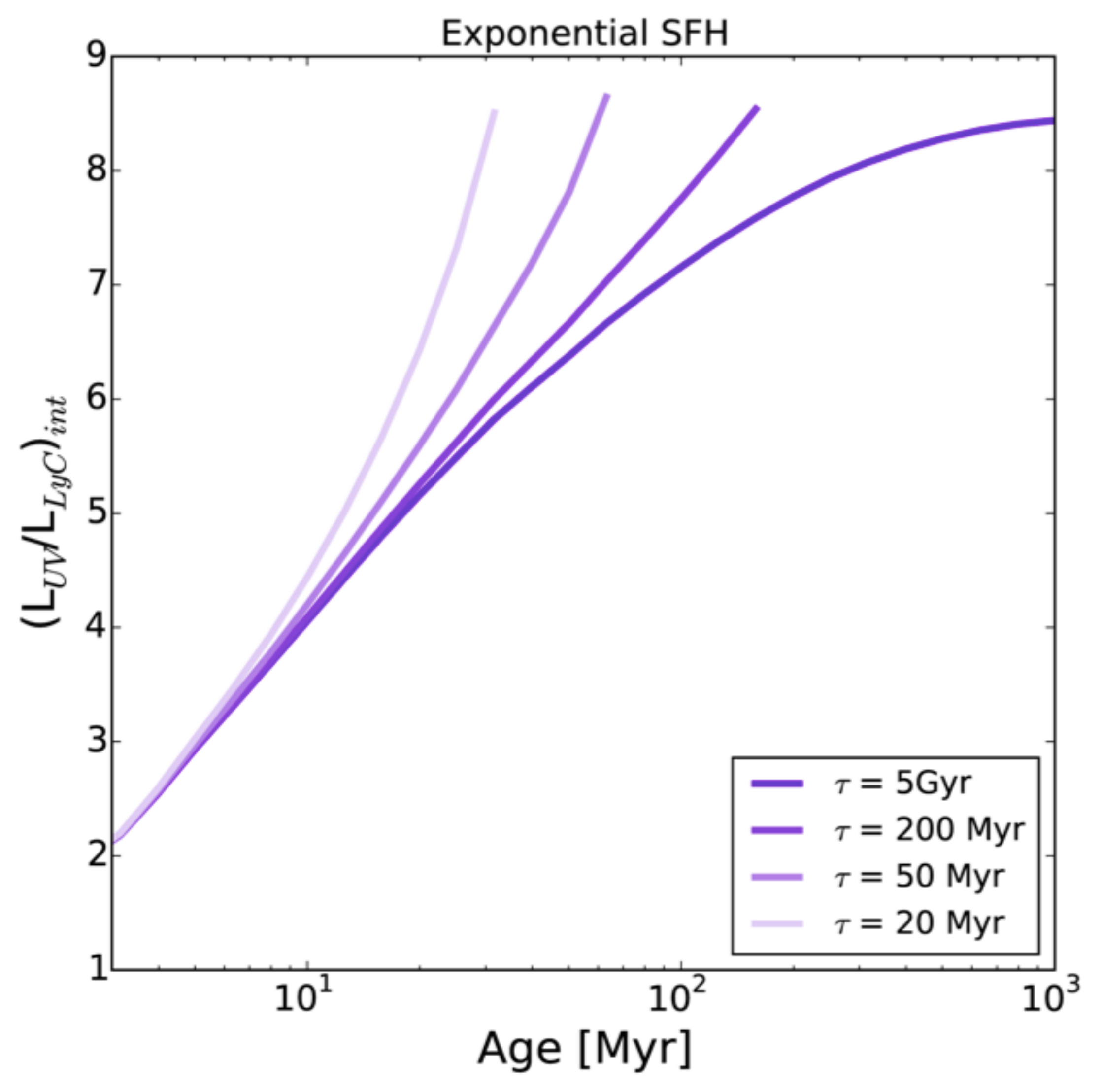}

\caption{The variation of the intrinsic UV--to--LyC luminosity ratio as
a function of the age of the stellar population and star--formation
history. In the upper figure, the ratios are derived from models
assuming constant star formation and varying metallicity. Here, we
reproduce (blue) the intrinsic ratio from {\it galexev} models with
constant star formation, allowing the metallicity of the model to vary
between 0.02 and 1$\times\!Z_{\odot}$.  Additionally, we produce models
using {\it Starburst99} for constant star formation history assuming a
constant solar metallicity.  In yellow, we illustrate the intrinsic
ratio for models in which the rotation speed of massive, main sequence
stars equals to zero and 40\% of the break-speed.  Furthermore, in green
we illustrate the intrinsic ratio as a function of age for models with
and without rotation assuming a top-heavy ($\alpha=-1.7$) IMF. In the
lower figure, the intrinsic ratio is derived exclusively from {\it
galexev} models assuming a range of exponentially declining SFHs.}
\label{fig:ratio} \end{center} \end{figure}

We consider the following models produced with {\it galexev}: 30
logarithmically-spaced ages between 10$^6$---$8\times10^8$ yrs, constant
and exponentially declining SFHs ($\tau$=0.02, 0.05, 0.2, 0.5, 1.0),
Salpeter (1955) and Chabrier (2003) initial mass function (IMF),  5
metallicities ($Z= 0.005, 0.02, 0.04, 1, 2.5\times\!Z_{\odot}$), and
Padova94 stellar evolution tracks.  For the composite models generated
with {\it Starburst99} we produce models with the same ages, constant
star formation histories, solar metallicity, Salpeter IMF and both
Geneva and Ekstr{\"o}m et al. (2012) evolutionary tracks. The latter
tracks include a detailed treatment of stellar rotation effects, and our
{\it Starburst99} library includes models with rotation speeds equal to
no rotation and rotation speed equal to 40\% of the break-up speed for
stars on the zero-age main sequence. Furthermore, we include models with
a top-heavy IMF slope ($\alpha=1.7$ at M$_{*}>0.5$\Msol).  For each
model in the library we computed ($L_{UV}/L_{LyC}$)$_{int}$, and the
results are summarized, as a function of the age of the stellar
population, in Figure~\ref{fig:ratio}.

In Figure~\ref{fig:ratio}, we only show exponentially declining SFH
models up to ages for which hot massive stars are still on the main
sequence (i.e., age$\,<\,\tau\,+\,25$~Myrs). This is because galaxies in our
sample were selected via the \ha\ emission line, implying the presence
of ionized gas. For a given star--formation history, and at any given
age, the stellar metallicity is found to have a minor impact on the
intrinsic ratio, introducing a scatter of the order of 20\%. The largest
uncertainty is introduced by our ignorance of the stellar age (more
specifically, the age of the latest burst of star formation) and the
IMF.  Additionally, Figure~\ref{fig:ratio} shows that the ratio can be
affected by stellar rotation, which reduces the intrinsic UV--to--LyC
ratio by approximately 20-30\% for a given IMF. Finally, we note that the choice of
stellar templates libraries ({\it galexev} vs. {\it Starburst99}) also affects the result, as shown by the comparison of the
solid dark blue and orange curves in Figure~\ref{fig:ratio}. 

Stellar population parameters can be derived using SED fitting
techniques (e.g., Skelton et al. 2014).  However, the resulting
parameters, particularly stellar age and SFR, are affected by large
systematic biases and uncertainties because of the assumed SFH are often
a poor representation of the real SFH of a galaxy (see, e.g., Dominguez
et al. 2015, Lee et al. 2009). Moreover, no constraints exist at high
redshifts on the IMF in star-forming regions.  For these reasons, and to
allow the reader to compare our results directly with other similar
studies, we adopt a conservative ($L_{UV}/L_{LyC}$)$_{int}=7$, noting
that the upper limits could be a factor of three {\it lower} smaller,
depending on the exact value assumed. 

\subsection{Absolute \fesc~from \ha\ luminosity} \label{sec:fescha}

The absolute escape fraction can also be computed directly (i.e.,
without first calculating $f_{esc,rel}$) if the intrinsic ionizing
emissivity can be estimated through indirect observations. By
definition, for each galaxy in our samples we have a measurement of its
rest-frame \ha\ luminosity, which can be used to estimate the amount of
ionizing radiation absorbed by neutral hydrogen in the galaxy. For a
luminosity--bounded HII region, the \ha\ luminosity produced by
recombination is simply proportional to the rate of ionizing photons,
i.e., $L_{\mbox{\tiny H$\alpha$}}\propto\,Q_{ion}$. 

In galaxies, however, the situation is more complex. First, dust is
known to be present inside HII regions, and absorbs $h\nu>13.6$eV
photons that would otherwise be able to ionize hydrogen, thereby
reducing the luminosity of the recombination lines.  Defining $f_{ion}$
as the fraction of ionizing photons absorbed by neutral hydrogen and
($1-f_{ion}$) the fraction of photons absorbed by dust in the galaxy,
the intrinsic rate of ionizing photons produced in the galaxy can be
written as $Q^{int}_{ion}=f_{ion}Q^{int}_{ion} +
(1-f_{ion})Q^{int}_{ion}$. This equation assumes that the fraction of
ionizing photons escaping from the galaxy is negligible, an assumption
justified by the result in the previous section. The first term in the
the expression for  $Q^{int}_{ion}$ will result in hydrogen ionization
and \ha-emission by recombination.  The second term will result in
thermal heating of the dust, and infrared re--emission, which will
depend on the dust grain properties and distribution of the dust with
respect to the ionizing sources.  No constraints exist on $f_{ion}$ in
high redshift galaxies.  In the local universe, however,  Inoue et al.
(2002) measured $f_{ion}$ in a sample of Galactic HII regions, using
infrared and radio  continuum observations. Because our observations
measure the \ha\ luminosity integrated over all  HII regions in the
galaxy, we will assume that $f_{ion}=0.5$, i.e.,  the average value
measured by Inoue et al. (2002). With this assumption, we can compute
the intrinsic rate of ionizing photons from the extinction--corrected
\ha\ luminosity.  

First, we correct the observed \ha\ luminosity for the contribution to
NII, by applying an average correction that depends on the galaxy's
stellar mass following Erb et al. (2006).  To correct the \ha\
luminosity for dust extinction we used a mass-dependent correction
defined in Dominguez et al. (2014) such that
$A_{H\alpha}$=3.3$\times$E(B-V)$_n$, where the color excess here equals
to the nebular color excess computed from stellar color excess as
$E(B-V)_n=2.27\times\,E(B-V)_s$ (see, e.g., Calzetti et al. 2000). We
use the best--fit $E(B-V)_s$ provided by Skelton et al. (2014). Assuming
the Calzetti extinction law ($\tau_{\lambda}$, Calzetti 1997), the
extinction--corrected \ha\ luminosity is then $L^{int}_{\mbox{\tiny
H$\alpha$}} = L_{\mbox{\tiny H$\alpha$}} \times10^{0.4E(B-V)_n
k_{\tau_{\mbox{\tiny H$\alpha$}}}}$. We then obtain the intrinsic rate
of ionizing photons from the unobscured \ha\ luminosity  as
$Q^{int}_{ion}=L_{\mbox{\tiny H$\alpha$}}^{int}
f^{-1}_{ion}c^{-1}_{\mbox{\tiny H$\alpha$}}$, (where $c_{\mbox{\tiny
H$\alpha$}}=1.37\times 10^{-12}$ erg, see e.g., Schaerer 2003). Finally,
we compute the intrinsic ionizing power from the rate of ionizing
photons as $\nu L_{LyC,int}= \epsilon_{ion} Q^{int}_{ion}$, where
$\epsilon_{ion}$ is the average energy per ionizing photon computed from
the same Starbust99 model used in the previous section\footnote{The
average energy per ionizing photon is computed from the {\it
Starburst99} template ($L(\nu)$) spectrum as
$\epsilon_{ion}=\frac{\int_{\nu_c}^{\infty}L(\nu)\rm{d}\nu}{\int_{\nu_c}
^{\infty}\frac{L(\nu)}{h\nu}\rm{d}\nu}$}, and $\epsilon_{ion}=16.2$eV
($\lambda=765$\AA). 

\begin{figure*}
\tiny
\begin{longtable}{ccccc} 
\caption{Emissivity-weighted \fesc~Upper Limits}\\ \hline \hline 
\multicolumn{1}{l}{\multirow{2}{*}{Redshift}} &
\multicolumn{1}{c}{\multirow{2}{*}{$\alpha$}} &
\multicolumn{1}{c}{} & 
\multicolumn{2}{c}{$\langle\!f_{esc}\!\rangle$} \\ \cline{4-5} \noalign{\vskip 2pt} 
\multicolumn{1}{c}{} & 
\multicolumn{1}{c}{} & 
\multicolumn{1}{c}{} & 
\multicolumn{1}{c}{$M_{UV,lim}=-13$} & 
\multicolumn{1}{c}{$L_{lim}=0.01\times\!L^*$}\\ \hline \hline \noalign{\vskip 2pt} \vspace*{-5pt}
\endhead
\multicolumn{5}{l}{{\sc Notes:} In Section 7, we measured upper limits to \fesc}\\
\multicolumn{5}{l}{\,\,\,for stacks of Grade A SFGs, sorted and binned according to}\\
\multicolumn{5}{l}{\,\,\,$M_{UV}$. We apply these}\\
\multicolumn{5}{l}{\,\,\,limits to measure an {\it emissivity--weighted} escape fraction, $\langle\!f_{esc}\!\rangle$,}\\
\multicolumn{5}{l}{\,\,\,from three sets of SFGs, with bin widths equal to}\\
\multicolumn{5}{l}{\,\,\,$\{-15\!\!\le\!\!M_1\!\!<\!\!-17.7;-17.7\!\!\le\!\!M_2\!\!<\!\!-18.4;-18.4\!\!\le\!\!M_3\!\!<\!\!-20.5$}\\
\multicolumn{5}{l}{\,\,\,We calculate $\langle\!f_{esc}\!\rangle$ at $z=1$ and $z=7$ assuming}\\
\multicolumn{5}{l}{\,\,\,Schechter parameters of the UV-luminosity function$:$}\\
\multicolumn{5}{l}{\,\,\,From Dahlen et al.\,(2007), $\{M^*,\phi^*\}=\{-19.9,3.99\times10^{-3}\}$}\\
\multicolumn{5}{l}{\,\,\,\,\,\,\,\,\,\,\,\,\,\,\,\,\,\,and $\alpha=-1.5,-1.9$}\\
\multicolumn{5}{l}{\,\,\,From Finkelstein et al.\,(2015), $\{M^*,\phi^*\}=\{-21.0,1.86\times10^{-4}\}$,}\\
\multicolumn{5}{l}{\,\,\,\,\,\,\,\,\,\,\,\,\,\,\,\,\,\,and $\alpha=-2.0$}\\
\endlastfoot
\multirow{2}{*}{$\,\,z=1$} & $-1.5$ & & 3.6\% & 3.4\% \\
& $-1.9$ & & 4.6\% & 4.3\%  \\ \vspace*{+2pt}
$z=7$& $-2.0$ & & 2.7\% & 2.4\% \\ \hline \hline \vspace*{+2pt}
\label{tab:longertab}
\end{longtable} 
\end{figure*}

For each group of galaxies, we compute the 3$\sigma$ limits on the
absolute escape fraction\footnote{To distinguish this value from the
absolute escape fraction obtained using $f_{esc,rel}$ (see
Section~6) we indicate it as $f_{esc}^{\mbox{\tiny H$\alpha$}}$.},
defined here as the ratio of the observed (corrected for IGM absorption)
to intrinsic ionizing power:

\begin{equation}
f_{esc}^{\mbox{\tiny H}\alpha}=\frac{\nu_{\mbox{\tiny eff}}L_{LyC,obs}\exp[\tau_{IGM}]}{\nu L_{LyC,int}},
\end{equation} \vspace{+5pt}

\noindent where $\nu_{\mbox{\tiny eff}}$ is computed from the pivot wavelength of the
FUV filter, divided by the median redshift of each galaxy sample
($\langle\,z\,\rangle = 1.1$, for all sub-samples). The results,
reported in the last column of Table~\ref{tab:longtab1}, are in good
agreement with the upper limits to \fesc~measured from the UV--to--LyC
continuum ratio. For example, for the Grade A sample \fesc$<$2.1\%, whereas
$f_{esc}^{\mbox{\tiny H$\alpha$}}<$3.2\%.

\section{Emissivity-weighted \fesc}

Theoretical models of  reionization assume a constant value of the absolute 
escape fraction as a function of galaxy UV luminosity. Here, we compute an 
emissivity--weighted \fesc~limit, that can be used to predict the number 
of ionizing photons for an observed non--ionizing UV luminosity function.
We define the emissivity--weighted \fesc~limit as:

\begin{equation} \langle f_{esc} \rangle=
\frac{\int_{L_{min}}^{L_{max}}{f_{esc}\cdot\!L\cdot\phi\left(\mbox{L}\right)}\mbox{dL}}{\int_{L_{min}}^{L_{max}}
{L\cdot\phi\left(\mbox{L}\right)}\mbox{dL}}, 
\end{equation} \vspace{+5pt}
\noindent where $\phi$(L) is the Schechter luminosity function (LF). 
To measure $\langle f_{esc}\rangle$, we first computed the upper limits
on \fesc~in three equally--sized Grade A SFG samples selected on
luminosity$:$ $-15.0>M_{1,UV}>-17.8$, $-17.8>M_{2,UV}>18.5$, and
$-18.5>M_{3,UV}>-20.5$.  Here, we include \fesc~in the integrand to
indicate that it will varies as a function of UV luminosity, though {\it
within} each of the luminosity bins it is assumed constant.  For these
three magnitudes ranges, applying the same technique as in Section 7.1,
we measure \fesc$<{5.6,3.9,1.7}\,[\%]$ respectively (3$\sigma$; see
Table \ref{tab:longertab}). 
 
We compute the integral in Equation 6 down to the faintest observed
magnitude ($M_{UV}=-15$, which corresponds to $\sim$0.01$L^*$ at $z=1$).
To quantify the contribution of the faintest sources, we also extend the
limit of integration down to $M_{UV}=-13$ (0.001$L^*$), assuming that
the $\langle f_{esc}\rangle$ limit remains constant down to the chosen
limit.  The resulting values of $\langle f_{esc}\rangle$ limits are
reported in Table \ref{tab:longertab} for $z\sim1$ and $z\sim7$,
assuming parameters for the luminosity function measured by Dahlen et al. (2007)
and Finkelstein et al. (2015), respectively. The emissivity weighted
\fesc~limits presented in Table \ref{tab:longertab} imply that the
observed population of SFGs contribute, on average, less than 50\% of
the ionizing background at $z\sim 1$.

We note here that the reported upper limits are valid for populations of
galaxies, and cannot be used for individual sources. In fact, in
calculating \fesc\ we are implicitly assuming that the  LyC is radiated
away from the galaxy isotropically. The orientation of a galaxy towards
the observer's line of sight may affect the identification of LyC
---starburst galaxies have been identified with strong winds and
outflows which could produce channels through which LyC may ``stream''
to the IGM (Heckman et al. 2001), though it is useful to note that
starburst galaxies with strong winds have not been detected to emit LyC
(Grimes et al 2009). In simulations as well, orientation biases in the
identification of LyC candidates are possible. Recently, Kimm and Cen
(2015) investigated, via cosmological hydrodynamical simulations, the
uncertainty introduced in the measurement of \fesc~ by stacking SFGs in
which LyC only escapes through channels with narrow opening angles. In
this work, we have stacked samples of {\it hundreds} of galaxies to
derive the upper limits; Kimm and Cen show that for such large samples,
the {\it uncertainty} on \fesc\,is $\lesssim20\%$.

\section{Discussion} 

Previous reports of the non-detection of escaping LyC in deep imaging of
$z\sim1$ SFGs, which necessarily produce LyC photons, imply that some
physical process(es) prevents this ionizing emission from escaping to
the IGM. In the previous section, we considered in a novel way the
escape of LyC from groups of galaxies defined with broadly similar
physical characteristics (e.g., morphology, star formation history) as
measured from multi-wavelength UV-optical, high spatial resolution
imaging and grism spectroscopy.  A detection of escaping LyC from
galaxies in any of these groups could indicate that particular physical
process(es) are more likely to promote the escape of LyC from SFGs. However, for all
all groups we measure only {\it upper limits} to \fesc. 

The direct detection of the LyC escape fraction in $z\sim1$ SFGs is not
essential to explain reionization at low redshift, because in this
redshift range, SFGs are the sub-dominant contributor to the ionizing UV
background. The upper limits to \fesc~presented here are more useful for
constraining the contribution of the {\it analogs} of SFGs at redshift
$z>4$. Although there are no constraints on the characteristics (e.g.,
mass, luminosity, metallicity) of LyC emitters at $z>4$, observations
and simulations suggest that the selection criteria applied in Section 4
should identify low-redshift analogs to the sources of reionization. For
example, the low-luminosity and low-mass galaxy sub-samples are measured
to have a mean $M_{UV}\simeq-16.5$. At $z\sim7$, this is equivalent to
$L\simeq0.02L^*$. These and lower-luminosity dwarf galaxies are inferred
form high redshift surveys to be critical sources of re-ionizing photons
(e.g., Robertson et al.~2015). Here, we discuss the implications of
these measured upper limits for reionization of the IGM at high redshift
($z\sim7$). 


At high redshift, the high column density of neutral absorbers prevents
the direct detection of LyC. Thus, the contribution of high redshift
SFGs to reionization is inferred from the observable non-ionizing UV
luminosity density ($\rho_{UV}$).  Specifically, $\rho_{UV}$ can be
related to the escape of ionizing photons following Madau, Haardt, and
Rees (1999), which Finkelstein et al. (2012) update as$:$
\begin{center}
\begin{equation} 
\begin{aligned}
 \rho_{UV} = {} & 1.24\times10^{25}\cdot\epsilon_{53}^{-1}\left(\frac{1+z}{8}\right)^3\left(\frac{\Omega_b\,h^{2}_{70}}{0.0461}\right)^2 \\ 
& x_{\mbox{\tiny \ion{H\!\!}{2}},z}\cdot\frac{C}{f_{esc}} [{\rm
  erg}\,{\rm s}^{-1}\,{\rm Hz}^{-1}\,{\rm Mpc}^{-3}] \end{aligned}
\label{eq:puveq}
\end{equation}
\end{center}\vspace{+5pt}

\noindent where $\epsilon_{53}$ is the number of Lyman continuum photons
per unit of forming stellar mass, in units of 10$^{53}$ photons
s$^{-1}$, $x_{\mbox{\tiny \ion{H\!\!}{2}},z}$ is the volume averaged
fraction of ionized hydrogen at redshift $z$, and $C$ is the ionized
hydrogen gas clumping factor)\footnote{A clumping factor equal to unity
corresponds to a homogenous medium.  In simulations, the range of
clumping factors typically assumed equals 1\lsim\,C\lsim\,50 (Gnedin and
Ostriker 1997; Pawlik, Schaye, \& van Scherpenzeel 2009; Finlator et al.
2012).} in the IGM ($C=\langle n^2\rangle/\langle n\rangle^2$). Here,
the prefactor results from equilibrating the volume averaged rate at
which ionized photons from young, hot stars are are emitted into the IGM
with the recombination rate of neutral gas in the IGM.  

With reasonable assumptions to $x_{\mbox{\tiny \ion{H\!\!}{2}},z}$ and
the clumping factor, we calculate the theoretical UV luminosity density
for comparison with high redshift observations. In this calculation, we
will assume \fesc\ at z=7 equals to $\langle$\fesc$\rangle<$2.7\%, which
is the upper limit we measured in Section 8 for $z\sim1$ SFGs integrated
over a comparable range of luminosity to allow for a direct comparison
with calculations of $\rho_{UV}$ measured at high redshift (e.g.,
Robertson et al.~2015, Finkelstein et al.~2015). We discuss the validity
of this assumption following the calculation. Assuming a clumping factor
typical for the high-redshift universe, C$\sim3$,
$\epsilon_{53}\simeq1$, and \fesc$<$2.7\%, then to {\it critically}
reionize the IGM at $z\sim$7 (\ifrac=1) requires
$\rho_{UV}\gtrsim1.3\times10^{27}\,{\rm erg}\,{\rm s}^{-1}\,{\rm
Hz}^{-1}\,{\rm Mpc}^{-3}$.  From direct observations reported in
Finkelstein et al. (2015), integrating the UV luminosity function
measured at $z\sim7$ over the range M$_{UV}$=[$-13,-23$],
$\rho_{UV}\sim1.3\times10^{26}\,{\rm erg}\,{\rm s}^{-1}\,{\rm
Hz}^{-1}\,{\rm Mpc}^{-3}$.  Thus, SFGs fail to meet the threshold for
critical reionization at $z\sim7$ by an {\it order of magnitude}. 

Implicit in this calculation are two noteworthy assumptions. First, we
assume that \fesc$_{,z=1}$=\fesc$_{,z=7}$. Multiple groups have measured
a strong positive evolution of the escape fraction of Ly$\alpha$ with
increasing redshift (Hayes et al.~2011; Kuhlen and Faucher-Gigu\'{e}re
et al.~2012). It is reasonable to expect the escape of LyC photons to be
proportional to that of Ly$\alpha$ photons (e.g., Verhamme et al. 2014)
therefore we may expect the escape of LyC photons may increase with
redshift. Secondly, the neutral fraction at $z\sim7$ may in fact be
non-zero, i.e., reionization is not fully complete until lower redshift
($z\lesssim6$).  Observations of a rapid evolution of Lyman-$\alpha$
emitters between $z\sim6-7$ suggest $0.2\lesssim\,x_{\mbox{\tiny
\ion{H\!\!}{2}},z}\lesssim0.6$ (see Pentericci et al. 2011). Thus, if
the escape fraction evolves with redshift or the volume averaged
fraction of ionized hydrogen is lower than was assumed here, the
disparity between the observed and predicted UV luminosity density would
be reduced and SFGs brighter than $M_{UV}<-13$ could in
principle maintain reionization at $z\sim7$.

It is interesting to compare what an \fesc=2.7\% would imply for
$\tau_{es}$, the Thomson scattering optical depth of CMB photons by free
electrons along the line of the sight (see, e.g., Nolta et al. 2009).
New results from the Planck Collaboration have revised the value of
$\tau_{es}$ downward to $\tau_{es}=0.063\pm0.012$ (Planck Collaboration
2015). Robertson et al. (2015) recently demonstrated that this revised
value of $\tau_{es}$ can be reproduced assuming \fesc$\sim 20\%$ and
integrating the contribution of SFGs as faint as $L=0.001L^*$, while
Finkelstein et al. (2015) also found $\tau_{es}$ remains consistent with
the Planck measurement even if \fesc~assumed for such faint SFGs is as
low as \fesc=13\%. Using $\rho_{UV}(z)$ calculated and compiled by
Finkelstein et al. (2015) for SFGs brighter than a limiting luminosity
$L=0.001L^*$, and assuming a {\it constant} \fesc$<$2.7\% over the
redshift range $4<z<11$ and $C=3$, we can infer an ionization history
over this redshift range directly from Equation \ref{eq:puveq}. From this
redshift-dependent ionization history, we compute an upper limit to
$\tau_{es}<0.059$ (3$\sigma$). This $\tau_{es}$ remains marginally
consistent (at $1\sigma$) with the latest measurement of $\tau_{es}$ by
the Planck Collaboration.

Finally, we note that the pressure to accomodate reionization at high redshift {\it
exclusively} with emission from SFGs has been recently called into
question.  Giallongo et al. (2015) recently identified 22 low-luminosity
$4<z<6$ AGN in the GOODS--S field --- 35\% of which were not previously
reported --- and suggested that the ionizing continuum from such objects
{\it alone} provides a dominant contribution to the reionization at high
redshift. Similarly, Madau and Haardt (2015) reconsider the ionization
history of hydrogen and helium in an AGN--dominated scenario and predict
$\tau_{es}=0.055$, which is similar to our upper limit derived for
$\tau_{es}$ derived for a reionization history dominated by stellar
sources of the hydrogen--ionizing emission in Figure~6.  In contrast,
Haardt and Salvaterra (2015) measure $\tau_{es}$ to be a factor of 2
{\it larger} than the recent Planck result when allowing for a larger
population of obscured AGN than Madau and Haardt (2015). Regardless,
considering the fact that we do not detect escaping LyC emission in the
stacked photometry of $\sim10^3$ SFGs, this AGN--dominated scenario is
appealing and in future work we will directly address this question
using these GALEX and HST data. 

\section{Conclusion}

Combining publicly available multi--wavelength archival data from
multiple deep imaging and spectroscopic surveys, we have made a renewed
search for escaping Lyman continuum emission in star--forming galaxies
at $z\sim1$. We select star--forming galaxies from catalogs of emission
line galaxies (ELGs) prepared from an independent reduction of HST WFC3
imaging and G141 grism data obtained originally as part of the 3DHST and
AGHAST suveys. Initially, we produced a large (N$\sim$1400) sample of
SFGs at \rsrang\, identified as such by the presence of H$\alpha$ in
emission. We apply a two--tiered selection against galaxies in this
initial sample to strongly exclude SFGs that could potentially suffer
photometric contamination from unresolved sources in the lower spatial
resolution GALEX images.  The final sample (``Grade A'') contained
$\sim$600 SFGs, and was used primarily for all analysis.  We find no
unambiguous evidence for escaping LyC in individual galaxies. Thus, we
stacked all galaxies in each sample and, using the optical photometry
compiled by the 3DHST team for the fields, derived 3$\sigma$ upper
limits to \fesc~ equal $\sim2$\%. In particular, we sub--divided the
Grade A sample on the basis of SFG's stellar mass, inclination and
H$\alpha$ equivalent width and report their associated upper limits.  In
particular, we measured an upper limit of \fesc\lsim9.6\% for ``extreme
emission line'' (H$\alpha>200$\AA) galaxies.

In addition, we have used an emissivity--weighted upper limit,
$\langle\!f_{esc}\!\rangle$, to constrain the contribution of SFGs to
the ionizing UV background near to the end of the reionization
($z\sim7$). This \fesc~upper limit was measured for $z\sim1$ SFGs, which
we assume to be reasonable analogs to the high redshift sources of
ionizing photons.  If the escape fraction of $z\sim7$ SFGs brighter than
M$_{UV}\sim-13$ is less than $\langle\!f_{esc}\!\rangle$$\sim$2.7\%, the
ionizing emissivity of these SFGs is insufficient to {\it critically}
reionize the high redshift universe.  If the LyC escape fraction
increases with redshift, SFGs remain plausible candidates for
reionization at high redshift and marginally consistent with results
from Planck, though we note that SFGs fainter than M$_{UV}>-17$ are not
observed in current surveys of the high redshift universe.
Alternatively, the contribution from an additional source (potentially
low--luminosity AGN) of ionizing photons could sustain reionization
during the Epoch of Reionization beginning at $z>11$ and completed by
$z\sim4-6$.

\acknowledgements

We thank referee, B. Robertson for helpful comments that improved the
discussion and conclusions presented in this work. We also thank S.
Finkelstein for helpful discussion.  M.H. acknowledges the support of
the Swedish Research Council (Vetenskapsr\r{a}det), the Swedish National
Space Board (SNSB), and the Knut and Alice Wallenberg Foundation. This
research was supported by NASA NNX13AI55G and HST--AR Program
\#12821.01, using observations taken by the 3D-HST Treasury program (GO
12177 \& 12328) with the NASA/ESA HST, which is operated by the
Association of Universities for Research in Astronomy, Inc., under NASA
contract NAS5-26555. GALEX and HST data presented in this paper were
obtained from the Mikulski Archive for Space Telescopes (MAST)
maintained by the STScI. Support for MAST for non--HST data is provided
by the NASA Office of Space Science via grant NNX13AC07G and by other
grants and contracts. This research has made use of NASA's Astrophysics
Data System Bibliographic Services. 


\appendix 

\begin{center} {\bf \large A. The Accuracy of $z_{spec}$ derived from
G141 Emission-Line Spectra} \end{center}

In general, we find good agreement between the photometric redshifts
reported in Skelton et al. (2014) and the grism spectroscopic redshifts
we used here --- $\Delta(z)=\frac{z_{spec}-z_{phot}}{1+z_{phot}} < 5\%$
for $\sim$50\% of all ELGs (2307/4481). 


We can improve the agreement between photometric and spectroscopic
redshifts to $\sim70\%$ by using the photometric redshift as a prior to
correct for mis-classified emission lines for which only a single
emission line was observed in the G141 spectrum. The inspection of a
spectrum obtained with a single grism is strongly susceptible to the
mis-classification of emission lines when only a {\it single line} is
observed.   By default, we classified such emission lines as H$\alpha$
and {\it not} a similarly bright, nebular emission line blue-ward of
H$\alpha$ ([OII$\lambda3727\mbox{\AA}$]
[OIII$\lambda\lambda4959,5007\mbox{\AA}$]), or other possibly prominent
lines red-ward of H$\alpha$ (HeI$\lambda10830\mbox{\AA}$, or the
recombination Paschen series).  Only in rare instances could the
morphology of the single emission line (i.e., a blue ``wing'' in an
unresolved [OIII] doublet) be used to overrule the default
classification, but unless both inspectors agreed to this classification
after second inspection these galaxies would not be included in our sample as we require both independent classifications of the emission line to agree for ELG selection.  Fortunately, extensive imaging of these
fields provides a long wavelength baseline for the measurement of
accurate photometric redshifts.  We apply the 3DHST collaboration's
public photometric redshift catalogs as a prior to constrain the
``true'' redshift for $\sim$800 ELGs (red circles of Figure 4) whose
spectra displayed only a single emission line.  This correction
marginally improves the size ($\sim$10\%) of the catalog of
\rsrang~SFGs for which the G141 is sensitive to the rest-frame
H$\alpha$ emission.  

Robust photometric redshifts are essential for the confirmation of
emission lines identified in grism surveys, but these measurements can
be incorrect as well. For a small subset of the galaxies (N$=$64), the
redshift discrepancy can be attributed to incorrect {\it photometric}
redshifts. In these SFGs, the emission line was catalogued as [OIII] by
inspection, implying a spectroscopic redshift $1.7\lesssim<z<2.2$, but
the photometric redshift is $\sim 30\%$ lower. An example is provided in
Figure \ref{fig:o3right}; this SFG clearly displays multiple strong
emission lines of H$\beta$ and [OIII], which is unresolved in the G141
grism. In Figure \ref{fig:o3right}, the 1D, 2D grism spectra, and the
F140W direct image of galaxy is provided; here, the direct image is
aligned with the orientation of the dispersion axis in the grism. Using
the correct spectroscopic redshift for these SFGs places them beyond the
range of interest, \rsrang, thus we excluded these galaxies from our
analysis. We include these galaxies in the public catalogs we have made
available online.

\begin{center}{\bf \large B. The sources of additional discrepancies
between $z_{spec}$ and $z_{phot}$}\end{center}

We observe a significant ($\Delta(z)>$5\%) discrepancy
between the 3DHST photometric redshift and the spectroscopic redshift
for a minority ($\sim$30\%) of ELGs.   This discrepancy can not be
corrected for using a photometric redshift prior as discussed in the
Appendix A. In this section, we discuss three likely scenarios that give
rise to these discrepancies.

 \begin{figure}[h!] \begin{center}
 \includegraphics[scale=0.25]{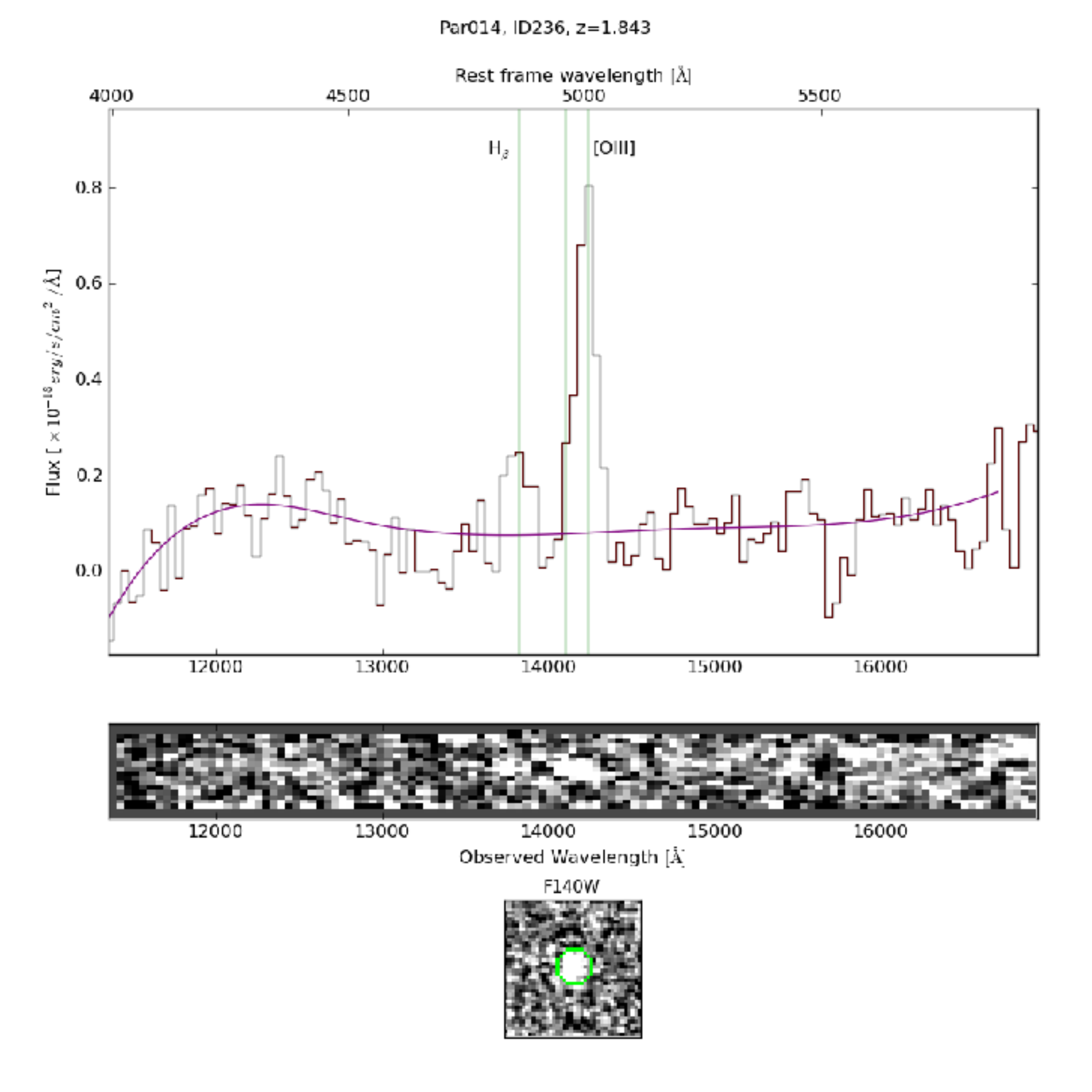}
 \caption{Representative galaxy ($<5$\% of all SFGs) for which the
 photometric redshift disagrees significantly with the measured
 spectroscopic redshift and can not be attributed to a mis-attribution
 of an emission-line(s).  Here, H$\beta$ and the unresolved [OIII] doublet are
 clearly identified in the one and two-dimensional spectra, implying the
 spectroscopic redshift is correct.} \label{fig:o3right} \end{center}
 \end{figure}

First, we investigate $\sim$150 $z_{spec}\simeq1.2$ single-emission line
SFGs whose discrepant redshifts could not be associated with a
mis-classified single emission ([OIII] was mis-classified as H$\alpha$)
in single emission-line spectra. Approximately 10\% of these galaxies
are observed to be spatially near to ($\ll$1 arcsecond), or blended
with, a neighboring galaxy. Automated photometry measurement software
will have some difficulty in distinguishing closely separated sources.
In some instances, the detection deblending parameters can be fine-tuned
to reduce source confusion (see, e.g., recent work with the UVUDF;
Rafelski et al. 2015) of close pairs or mergers in which the system can
be resolved as two distinct galaxies, but ultimately the photometry will
be contaminated in the close pair system.   Such galaxies are only a
small fraction of all ELGs, thus the flux blending due to source
confusion is likely to be only a minor factor in the redshift
discrepancy for the full sample.   

Secondly, and more importantly, we noted in visual
inspection that most of these galaxies ($\sim$70\%) are faint
(m$\lesssim$24) and compact (r$_e\sim$3 pixels) in the direct image.
This raises the possibility that the emission line, attributed as
H$\alpha$, could contribute significantly to the broad-band magnitude
(see, e.g., Atek et al. 2014) and thus affect the measurement of the
photometric redshift. In Figure \ref{fig:linecont} we measure the
contribution of the observed emission line to the observed magnitude in
the F140W filter. Here, a two-dimensional histogram of galaxies with
correctly-identified H$\alpha$ (i.e.,
$\frac{z_{spec}-z_{phot}}{1+z_{phot}}<$5\%)--- are overplotted. Red data
indicate those galaxies in the ``cluster'' which have discrepant
redshifts.

\begin{figure}[h!]
\begin{center}
\includegraphics[scale=0.35]{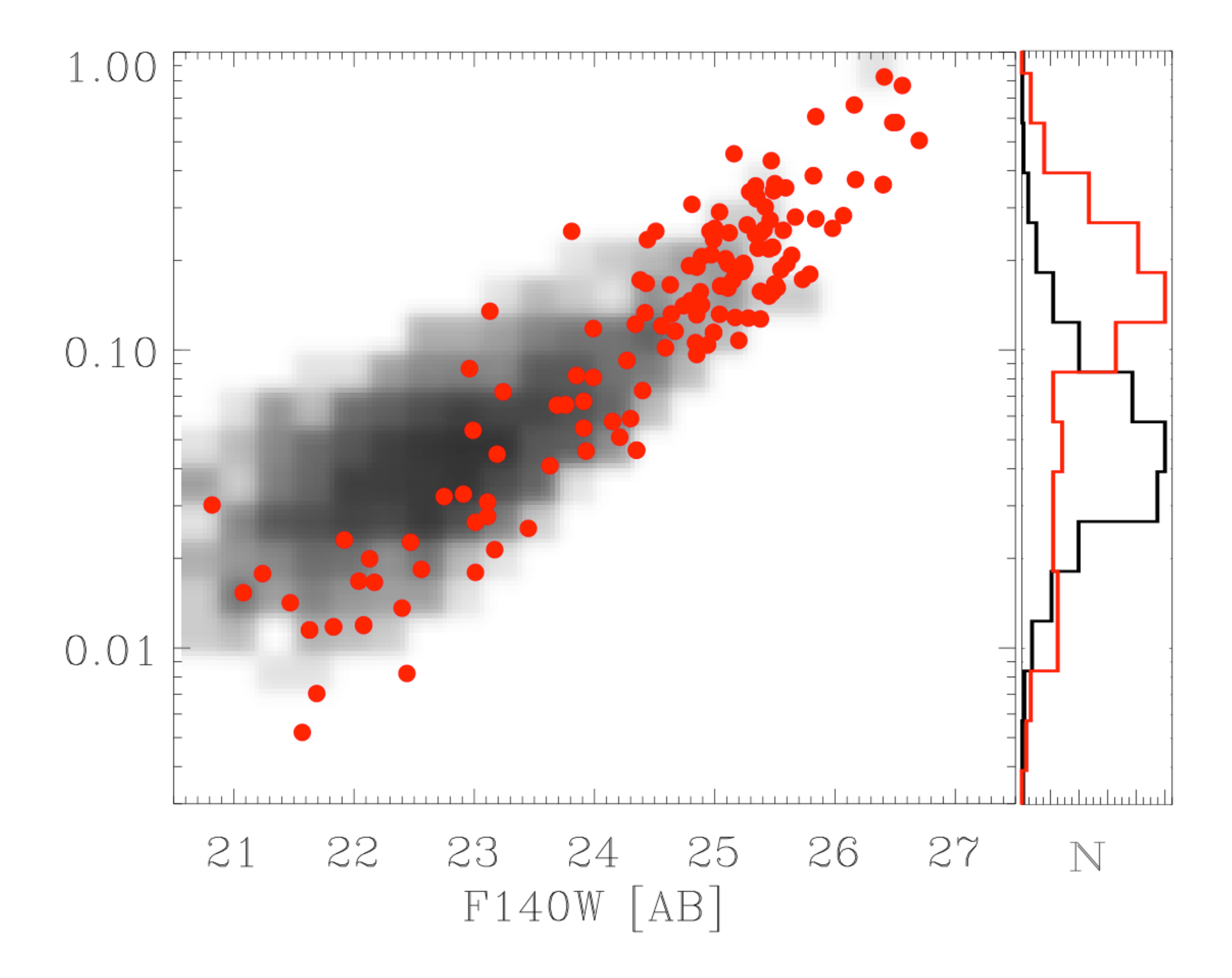} 

\caption{We measured the contribution of the H$\alpha$ emission line to
the broadband F140W photometry, presented here as a percentage of the
total flux, for two subsets of SFGs. Gray, binned data indicates SFGs
with grism spectroscopic redshifts in good agreement with published
3DHST photometric redshifts.  Red data indicate $z\sim2$ SFGs with
single emission lines and discrepant redshifts, but whose lines could
not be re-classified from H$\alpha$ to [OIII] using the photometric
redshift as prior on the classification. The two distributions are
clearly offset; here, the SFGs with discrepant photometric redshifts can
here be attributed to the contribution of strong emission from a high EW H$\alpha$.}

\label{fig:linecont} \end{center} \end{figure}

In Figure \ref{fig:linecont}, there is a clear distinction in the
centroids of the distribution of the gray-scale (all galaxies) and
discrete, red (strong contribution of H$\alpha$ to the continuum flux)
data. The contribution from line emission ($>$10\%, and extending to
nearly 100\%) to the broadband F140W magnitude can affect the SED of the
galaxy ($\sim$0.5 magnitudes Atek et al. 2011) and thus the inferred
physical characteristics \citep[see, e.g.,][]{deBarros14} and
photometric redshift.  If the IR photometry is affected by strong H$\alpha$, additional 
nebular lines may too be bright and the galaxy's broadband SED at shorter rest-frame wavelengths may also be adversely affected.  In the case where strong emission lines are observed in faint ELGs, we take a conservative approach
and choose to exclude these galaxies from the Grade A/B sample.  If the mis-classified emission line is
associated with a galaxy at lower ($z\lesssim0.9$) redshift, flux in the
GALEX FUV filter would be assqociated with the non-ionizing continuum
and these sources would contaminate the measurement of the \fesc.

Third, we consider those SFGs identified at $0.9<z_{spec}<1.4$, but
$0<z_{phot}<0.9$.  In this sub-sample, the H$\alpha$
 emission line contributes  $\gtrsim$10\% of the total flux in the
broadband filter in only $\sim$30\% of the galaxies, thus the accuracy
in the measured photometric redshift is not singularly at issue.
Instead, we attribute the redshift discrepancy measured for this subset
of galaxies to at least one of four additional issues. First, nearly
half of all sources were measured with an H$\alpha$ equivalent width
less than $40\mbox{\AA}$, making the positive identification by eye more
difficult in galaxies with fainter continuua.   HST grism surveys are
largely incomplete to such low EW sources (Colbert et al. 2013) and we
removed sources with such low EWs from our sample selection in Section
\ref{sec:selection} but discuss them here for completeness. Of the
remaining 55\% of sources with a redshift discrepancy $\Delta(z)>5\%$,
$\sim$50\% of these galaxies amongst the brightest studied
($m_{F140W}<23$) and their grism spectra were correspondingly dominated
by a very bright continuum spectral profile, which made it difficult to
visually confirm the presence of an emission line.  A visual inspection
of all galaxies direct image also confirms that approximately 50\% of
all galaxies were large spiral galaxies. These observations suggest that
the photometric redshifts are indeed accurate and the discrepancy in
redshifts is likely attributable to an incorrect spectroscopic
redshifts. We attempted to confirm the accuracy of the photometric
redshifts for these galaxies, assuming the candidate emission line was
associated with near-IR emission lines of HeI and the Paschen series but
these results were inconclusive as often only a single emission line was
identified in the grism spectra. Amongst the full sample of these
low-redshift discrepant ELGs, we attribute an additional 50\% (for a
total of $\sim$100\%, including the bright continuum and spiral
galaxies) of redshift discrepancies to contamination issues.   Sources
of contamination included 1) overlapping 1$^{st}$ (or higher) order
spectra in the extracted spectra from galaxies in pairs, groups, or in
close proximity to bright galaxies ($\sim60\%$) and 2) noise arising
from the proximity of the dispersed spectrum to either the dead pixel
region\footnote{the ``Death Star'', see HST IR Instrument Science Report
\#WFC3-2008-28} or near the  edge of the IR detector ($\sim$40\%). We
elect not to correct the redshifts for these galaxies --- if the
photometric redshift is incorrect, the GALEX FUV filter would be
sensitive to significant non-ionizing UV continuum which would again
contaminate the measurement of \fesc.



\end{document}